\documentclass[conference]{IEEEtran}
\usepackage{cite}
\usepackage{amsmath,amssymb,amsfonts}
\usepackage{algorithmic}
\usepackage{braket}
\usepackage{graphicx}
\usepackage{textcomp}
\usepackage[dvipsnames]{xcolor}
\usepackage{fancyhdr}
\usepackage{comment}
\usepackage[hyphens]{url}
\usepackage{adjustbox}
\newtheorem{theorem}{Property}
\usepackage{bm}
\usepackage[linewidth=1pt]{mdframed}

\newcommand\blfootnote[1]{%
  \begingroup
  \renewcommand\thefootnote{}\footnote{#1}%
  \addtocounter{footnote}{-1}%
  \endgroup
}

\usepackage{soul}
\usepackage{booktabs}
\usepackage[bookmarks=true,breaklinks=true,letterpaper=true,colorlinks,linkcolor=black,citecolor=blue,urlcolor=black]{hyperref}
\usepackage{adjustbox}

\usepackage[linewidth=1pt]{mdframed}

\usepackage{tikz}
\newcommand*\circled[1]{\tikz[baseline=(char.base)]{
            \node[shape=circle,draw,inner sep=0.5pt] (char) {#1};}}
\usepackage{mathtools}

\usepackage[font=bf]{caption}

\usepackage{tabularx}

\usepackage[caption=false]{subfig}

\newcommand{\subparagraph}{}
\usepackage[utf8x]{inputenc}
\usepackage{authblk}

\def\BibTeX{{\rm B\kern-.05em{\sc i\kern-.025em b}\kern-.08em
    T\kern-.1667em\lower.7ex\hbox{E}\kern-.125emX}}

\title{VAQEM: A Variational Approach to \\ Quantum Error Mitigation }
      
\author[1]{Gokul Subramanian Ravi*}
\author[1]{Kaitlin N. Smith*}
\author[2]{Pranav Gokhale}
\author[3]{Andrea Mari}
\author[4]{Nathan Earnest}
\author[4]{\\Ali Javadi-Abhari}
\author[1]{Frederic T. Chong}
\affil[1]{University of Chicago}
\affil[2]{Super.tech}
\affil[3]{Unitary Fund}
\affil[4]{IBM Quantum, IBM T.\ J.\ Watson Research Center}

\begin{document}

\maketitle
\pagestyle{plain}
\blfootnote{

\noindent * Both authors contributed equally. Correspondence: gravi@uchicago.edu and kns@uchicago.edu. 

\noindent To appear at The 28th IEEE International Symposium on High-Performance Computer Architecture (HPCA-28).}


\begin{abstract}

Variational Quantum  Algorithms (VQA) are one of the most promising candidates for near-term quantum advantage.
Traditionally, these algorithms are parameterized by rotational gate angles whose values are tuned over iterative execution on quantum machines. 
The iterative tuning of these gate angle parameters make VQAs more robust to a quantum machine’s noise profile.
However, the effect of noise is still a significant detriment to VQA's target estimations on real quantum machines - they are far from ideal. 
Thus, it is imperative to employ effective error mitigation strategies to improve the fidelity of these quantum algorithms on near-term machines.

While existing error mitigation techniques built from theory do provide substantial gains, the disconnect between theory and real machine execution characteristics limit the scope of these improvements. 
Thus, it is critical to optimize mitigation techniques to explicitly suit the target application as well as the noise characteristics of the target machine. 

We propose VAQEM, which dynamically tailors existing error mitigation techniques to the actual, dynamic noisy execution characteristics of VQAs on a target quantum machine. 
We do so by tuning specific features of these mitigation techniques similar to the traditional rotation angle parameters - by targeting improvements towards a specific objective function which represents the VQA problem at hand. 
In this paper, we target two types of error mitigation techniques which are suited to idle times in quantum circuits:  single qubit gate scheduling and the insertion of dynamical decoupling sequences. 
We gain substantial improvements to VQA objective measurements --- a mean of over 3x across a variety of VQA applications, run on IBM Quantum machines.

More importantly, while we study two specific error mitigation techniques, the proposed variational approach is general and can be extended to many other error mitigation techniques whose specific configurations are hard to select a priori.
Integrating more mitigation techniques into the VAQEM framework in the future can lead to further formidable gains, potentially realizing practically useful VQA benefits on today's noisy quantum machines.

\end{abstract}

\section{Introduction}

Quantum computing is a revolutionary computational model that leverages quantum mechanical phenomena for solving some classically-intractable problems. 
Quantum computers (QCs) evaluate quantum circuits or programs in a manner similar to a classical computer, but quantum information's ability to leverage superposition, interference, and entanglement is projected to give QCs significant advantage in chemistry~\cite{kandala2017hardware}, optimization~\cite{moll2018quantum}, and machine learning~\cite{biamonte2017quantum}.

In near-term quantum computing, sometimes called Noisy Intermediate-Scale Quantum (NISQ), we expect to work  with quantum machines which comprise hundreds to thousands of imperfect qubits ~\cite{preskill2018quantum}. 
Further, due to design constraints, the connectivity in these machines will be sparse and qubits will have modest lifetimes. 
With such limitations, near-term machines will be unable to execute large-scale quantum algorithms like Shor Factoring~\cite{Shor_1997} and Grover Search~\cite{Grover96afast}, which would require error correction comprised of millions of qubits to create fault-tolerant quantum systems~\cite{O_Gorman_2017} wherein many physical device qubits work in careful synchronization to create a single logical qubit that is used at the algorithm level.

\begin{figure}[t]
\centering
\includegraphics[width=\columnwidth,trim={0cm 0cm 0cm 0cm}]{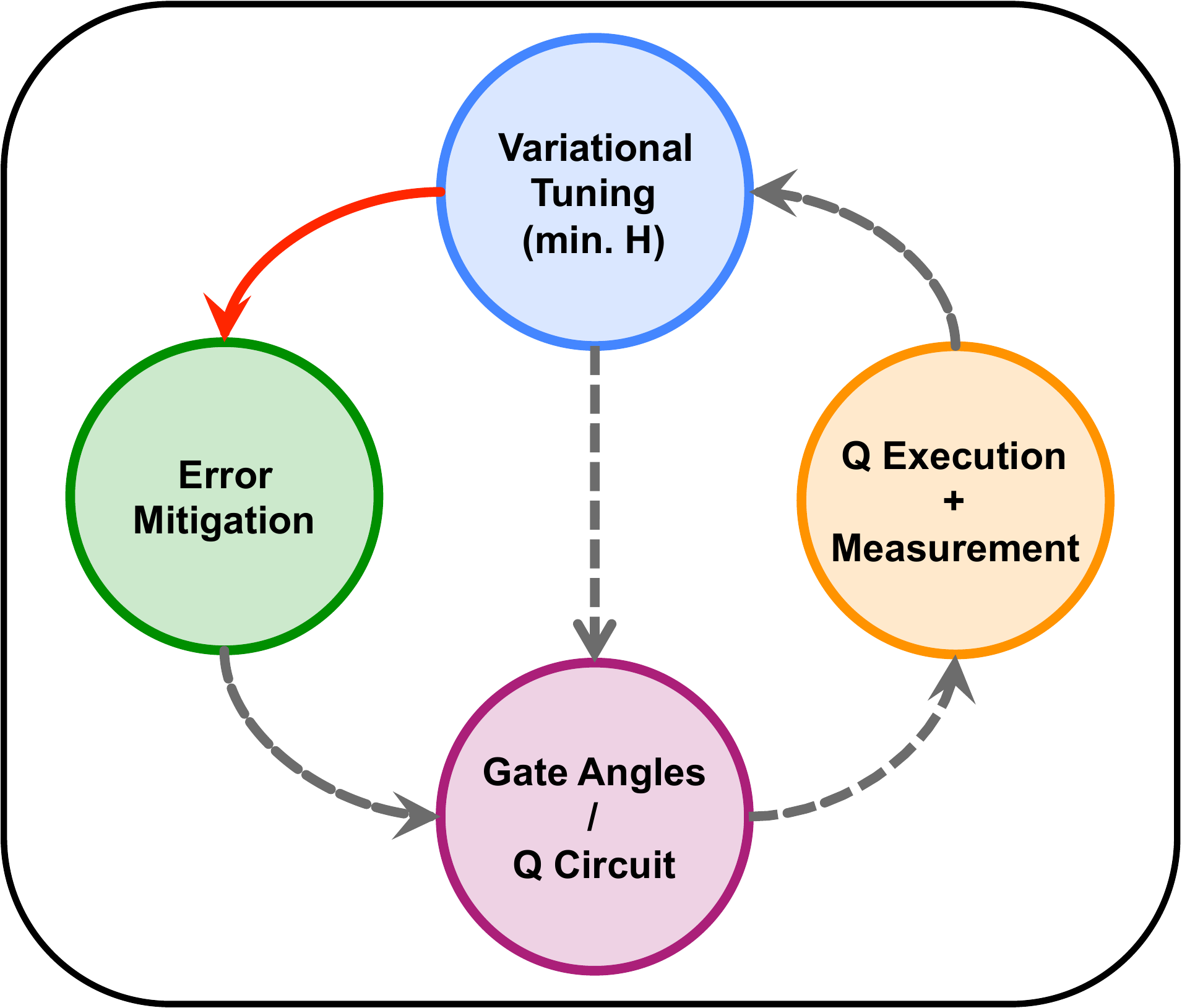}
\caption{In traditional VQA applications, the gate angles to the quantum circuit (ansatz) are tuned variationally to optimize for (eg. minimize) some objective function / Hamiltonian H. Further, error mitigation is often applied to the quantum circuit but is usually built form theory / heuristics and is independent of the tuning process. The entire traditional flow is indicated by the dashed grey arrows. In this work we propose to variationally improve the error mitigation techniques to suit the application / machine as well (red arrow). We tune specific features of the error mitigation techniques within the VQA harness of minimizing the same Hamiltonian H. This effectively brings VQA closer to the optimal noise-free solution.}
\label{fig:intro}
\end{figure}


\textbf{\emph{Variational Quantum Algorithms:}} However, recently, variational quantum algorithms (VQA) have been introduced that are expected to be a near-perfect match for near-term machines where logical and physical qubits have a one-to-one mapping. This class of algorithms has a wide range of applications such as ground state energy estimation of molecules~\cite{peruzzo2014variational}, MAXCUT approximation~\cite{moll2018quantum}, prime factorization~\cite{anschuetz2018variational} and so on. 
The fundamental aspect of VQAs that make them suitable to today's quantum devices is that these algorithms \emph{adapt} to the characteristics and noise profile of the quantum machine they are executing on.
The quantum circuit for a VQA is parameterized by a list of angles (which are the angles corresponding to gate rotations in the circuit). These parameters are optimized by a classical optimizer over the course of many iterations to maximize or minimize a specific target objective which is representative of the VQA problem. For this reason, variational algorithms are also termed as hybrid quantum-classical algorithms.
These VQA characteristics make them more robust to be executed on today's  quantum devices.

\textbf{\emph{Noise in near-term Machines:}} As the name suggests, NISQ devices suffer from high error rates as noise is prevalent during state initialization, gate application and measurement procedures.
In addition to errors incurred while directly interacting with qubits, quantum state is also vulnerable to error during periods of inactivity. This noise on qubits is termed decoherence and manifests as damping that degrades the quality of qubit state exponentially over time. 
Noise has prevented current quantum computers from surpassing the capabilities of classical computers in almost all applications, including VQAs.
Considering the promise of VQAs on NISQ machines, it is imperative to explore techniques to improve the quality of their execution on today's quantum machines.

\textbf{\emph{Error Mitigation:}} Orthogonal to the domain of VQAs, multiple error mitigation techniques have been explored on NISQ devices.
These techniques reduce the effect of noise on circuit execution on the quantum machine.
Several promising strategies have recently emerged, including noise-aware compilation~\cite{murali2019noise,tannu2019not}, scheduling for crosstalk~\cite{murali2020software,ding2020systematic}, 1-qubit gate scheduling in idle windows~\cite{smith2021error}, dynamical decoupling~\cite{viola1999dynamical,pokharel2018demonstration, souza2012robust}, zero-noise extrapolation~\cite{temme2017error,li2017efficient,giurgica2020digital}, readout error mitigation~\cite{tannu2019mitigating,bravyi2021mitigating}, exploiting quantum reversibility~\cite{patel2021qraft,smith2021error} and so forth.
While these techniques have the potential to greatly improve execution fidelity, there is often a significant disconnect between the theoretical guarantees of these methods and how they perform on real hardware. Complicated sources of error on actual hardware make it difficult to reason about the best configuration of the error mitigation technique.

\textbf{\emph{A Variational Approach to Error Mitigation:}} Dynamically tailoring specific features of error mitigation techniques to a machine's actual noisy execution environment is a perfect fit for the variational quantum approach which already tunes gate angle parameters as part of the framework. Thus, we propose VAQEM, a variational approach to error mitigation which integrates error mitigation features into VQA's iterative tuning of circuit parameters towards the target problem objective.
An overview of VAQEM is illustrated in Fig.\ref{fig:intro}.
In this paper, we target two types of error mitigation techniques which are suited to idle times in quantum circuits:  single qubit gate scheduling~\cite{smith2021error} and the insertion of dynamic decoupling sequences~\cite{viola1999dynamical,pokharel2018demonstration, souza2012robust}, resulting in substantial improvements in the measured objective over a variety of VQA applications.

Most importantly, while we limit our scope to two idle-time error mitigation techniques, the proposed variational approach can enable improved mitigation over a variety of other techniques, paving the way for quantum advantage for VQA in the NISQ era. 
Further, benefits of this new approach will only improve as the support for classical-quantum co-processing in the cloud continues to grow.

\textbf{\emph{Contributions:}}

\circled{1}\ VAQEM is the first variational approach to error mitigation which integrates error mitigation techniques into VQA's harness of feedback-based parameter tuning. It is able to do so without getting lost in the increased degrees of tuning freedom.

\circled{2}\ The benefits from the VAQEM approach are demonstrated by showing improvements in two idle-time error mitigation methods - insertion of dynamical decoupling  and single qubit gate scheduling. Each idle window has a different optimal error mitigation “configuration” which is dependent on input state to the  idle window, the qubit characteristics etc., and therefore it is not trivial to identify the optimal configuration.

\circled{3}\ To the best of our knowledge, VAQEM is the first proposal presenting the effective selection of dynamical decoupling (DD) sequence lengths tailored to a quantum device.

\circled{4} Our techniques are evaluated on real IBM quantum machines. VAQEM improves the quality of the measured objective by over 3x on average across a variety of VQA applications. 

\circled{5}\ To the best of our knowledge, this work is among the first research papers that use the IBM Qiskit Runtime\cite{qiskitruntime} framework, which allows for greater than 100x speedups in classical-quantum co-processing and is necessary to extend VQA research on real quantum machines. We also discuss the challenges of execution on today's quantum machines via the available cloud infrastructure and their effects on tackling VQA problems.

\circled{6}\ We prove the ``soundness" of the objective-aware feedback-based tuning approach that we utilize to reduce error from noise in quantum circuits.

\circled{7}\ We discuss the potential for VAQEM to be extended to other error mitigation and QC optimization approaches. Incorporating multiple techniques into  the variational framework always ensures synergistic improvements because any negative interactions between techniques are avoided by the tuning mechanism.

\section{Background}

\subsection{Near-term Quantum Computing}
\label{information-qc}

The basic unit of quantum information is the quantum bit, or qubit, which can exist in a linear superposition of the basis states $\ket{0}$ and $\ket{1}$. If measured, however, the state $\ket{\psi} = \alpha\ket{0} + \beta\ket{1}$ collapses into either $\ket{0}$ or $\ket{1}$, effectively becoming a classical bit. A system of $n$ qubits requires $2^n$ amplitudes to describe the state.

Qubits are manipulated via gates which modify their amplitudes. Unlike in classical computation, there are many non-trivial single-qubit gates such as $R_x(\theta)$ and $R_z(\theta)$ which rotate the state around the $x$ and $z$ axes, respectively. Pairs of qubits can be manipulated via multiqubit interactions. The most common of these gates is the two-qubit controlled-$X$, or $CX$. Together with single qubit gates, $CX$ enables universal quantum computation. There are many choices of basis gate sets specified by the underlying hardware.

Current QCs, sometimes called Noisy Intermediate Scale Quantum (NISQ) devices, are error prone and less than 100 qubits in size~\cite{preskill2018quantum}. These devices are extremely sensitive to external influence and require precise control, and as a result, some of the biggest challenges that limit scalability include limited coherence, gate errors, readout errors, and connectivity. 
Environmental coupling is the source of many errors in quantum systems. For example, amplitude damping describes the sporadic loss of energy resulting in the $\ket{1}$ state falling to the $\ket{0}$ state; the rate of this process is described by the device's $T_1$ time. Similarly dephasing details the sporadic change in relative phase between basis states and is expressed by the $T_2$ time of the device. Both are the cause of decoherence. Finally, crosstalk refers to error caused by simultaneous execution of gates on nearby qubits. The severity of each type of noise varies per qubit and calibration cycle.

\begin{figure}[!t]
\centering
\fbox{
\includegraphics[width=\columnwidth,trim={0cm 0cm 0cm 0cm}]{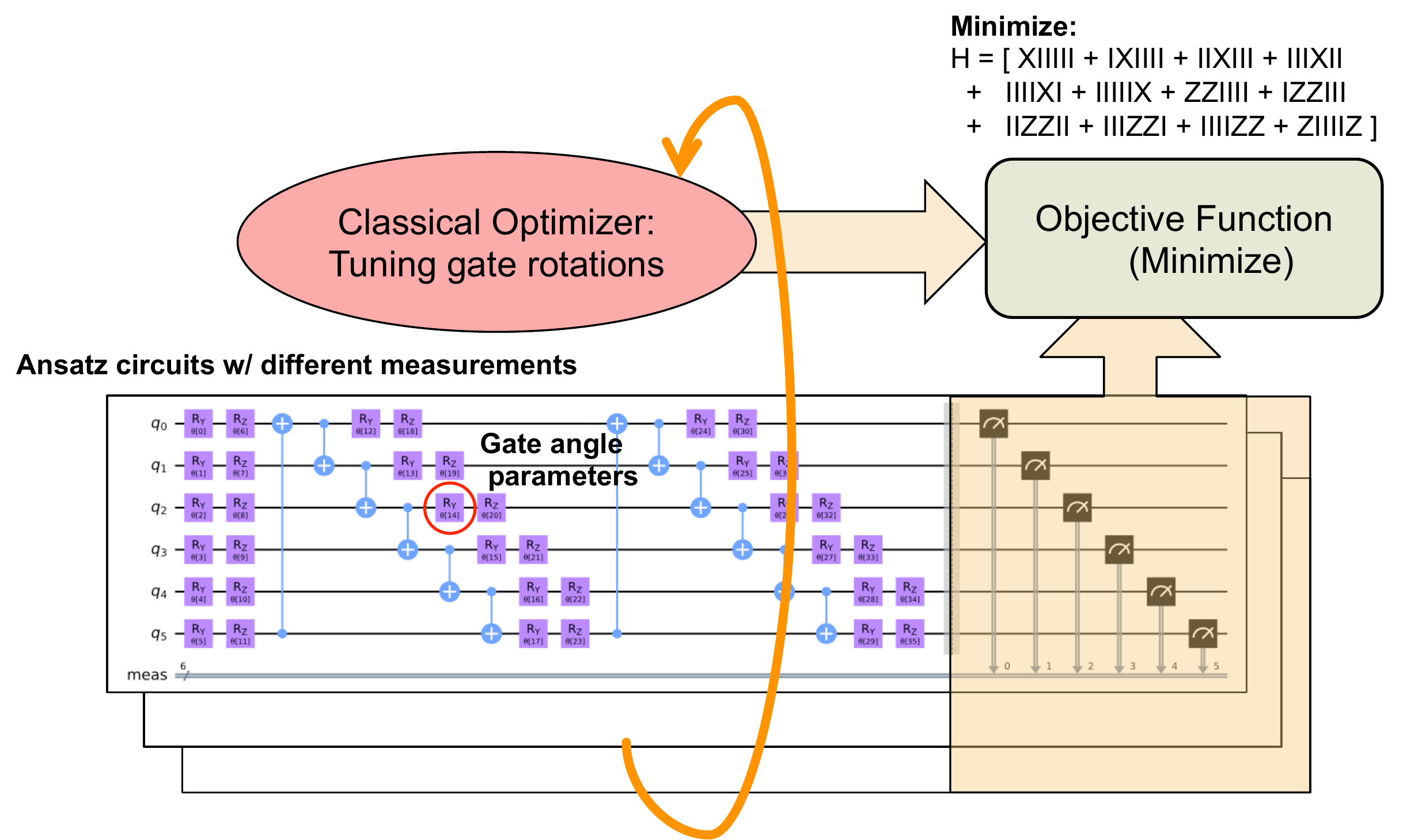}
}
\caption{Overview of VQA flow that includes quantum and classical processing to minimize some objective function $<H>$, the expectation value of the problem Hamiltonian.}
\label{fig:vqa}
\end{figure}

\subsection{Variational Quantum Algorithms}
\label{VQA}
Variational quantum algorithms are important in the near-term because they comply with the constraints of NISQ hardware.
In particular, variational algorithms have innate error resilience, due to the hybrid alternation with a noise-robust classical optimizer~\cite{peruzzo2014variational, mcclean2016theory}. 
A schematic of this process is illustrated in Fig.~\ref{fig:vqa}.
There are multiple applications which are part of the VQA domain such as Quantum Approximate Optimization Algorithm (QAOA)~\cite{farhi2014quantum} and Variational Quantum Eigensolver (VQE)~\cite{peruzzo2014variational} .
In this work we restrict ourselves to the latter, which is discussed next.

\subsubsection{Applications: Variational Quantum Eigensolver}
The Variational Quantum Eigensolver (VQE)~\cite{peruzzo2014variational} is used to find the lowest eigenvalue of a given problem matrix by computing a difficult cost function on the quantum machine and feeding this value into an optimization routine running on a CPU. Typically, the problem matrix is the Hamiltonian governing a target system and the lowest eigenvalue corresponds to the system's ground state energy~\cite{mcclean2016theory}.
VQE is typically used to find the ground state energy of a molecule, a task that is exponentially difficult in general for a classical computer~\cite{Gokhale:2019}.
Estimating the molecular ground state has important applications to chemistry such as determining reaction rates and molecular geometry. 

At a high level, VQE can be conceptualized as a guess-check-repeat algorithm. 
The check stage involves the preparation of a quantum state corresponding to the guess.
This preparation stage is done in polynomial time on a quantum computer, but would incur exponential cost  in general on a classical computer. 
This contrast gives rise to a potential quantum speedup for VQE~\cite{Gokhale:2019}.

\begin{figure}[!t]
\centering
\fbox{
\includegraphics[width=0.95\columnwidth,trim={1cm 3.5cm 0.7cm 3.5cm}]{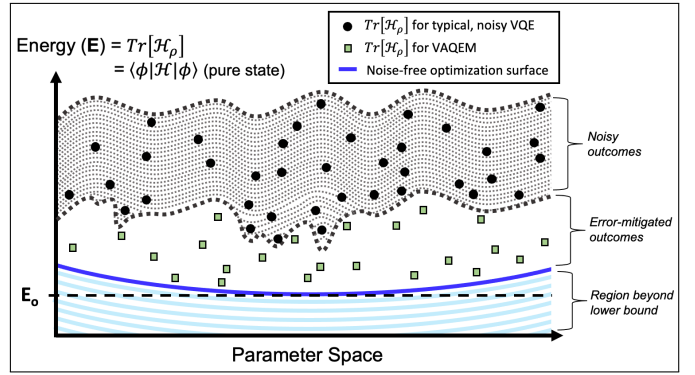}
}
\caption{Comparison of the ideal VQE optimization surface vs. local minimums ($Tr[\mathcal{H}_\rho]$) found in the traditional, noisy and error-mitigated search regions.}
\label{fig:vqa-search-space}
\end{figure}

\subsubsection{Ansatz}
The quantum circuit corresponding to each iteration of VQE (and VQA in general) is termed an ansatz which describes the range of valid physical systems and thus determines the optimization surface. 
Traditionally, the ansatz is parameterized by gate rotation angles as shown in Fig.~\ref{fig:vqa}.
A good ansatz provides a balance between a simple representation, efficient use of available native hardware gates, and sufficient sensitivity with the input parameters.

For molecules, while many ansatz choices are possible, Unitary Coupled Cluster Single-Double (UCCSD), an ansatz motivated by principles of quantum chemistry, is considered the gold standard~\cite{romero2018strategies, Gokhale:2019}. 
Unfortunately the UCCSD ansatz is generally of considerable circuit depth making it less suitable for today's NISQ machines except for very small molecules like H$_2$.
More suitable to the NISQ-era are hardware efficient ansatz~\cite{kandala2017hardware} like the SU2~\cite{IBM-SU2} which are hyper-parameterized by number of qubits, number of repetitions, and type of entanglement, and these can be selected so as to optimize for the quantum application as well as the machine.

\subsubsection{Hamiltonian}
The VQA problem is represented as a Hamiltonian.
The VQA objective function calculates the expectation value of this Hamiltonian iteratively.
This objective function is derived from ansatz measurements over different bases - an example is shown in Fig.~\ref{fig:vqa}.

For example, in VQE the classical tuner variationally changes the parameterized input until it converges to a global minimum.
This way it finds the corresponding eigenvalue and eigenstate. 
Since VQE's Hamiltonian describes the energy evolution, this global minimum represents the ground state energy of the system. QAOA takes a similar approach.

If a circuit ideally intends to simulate the evolution of some Hamiltonian, $\mathcal{H}$, then in a noisy environment it can be represented as the evolution of some other Hamiltonian, $\mathcal{H'}$. If noise levels are reasonable and/or if the effects of noise are adequately mitigated then the eigenstates and eigenvalues of $\mathcal{H}$ and $\mathcal{H'}$ will be close in distance.

\subsubsection{Parameter Space and Classical Tuning}
The execution on the quantum processor evaluates the objective function to be optimized classically. 
A good optimizer has a short distance to the true global optimum, high accuracy of the optimal parameters found, or both.~\cite{9259985} .
The shape of the optimization surface is determined by the ansatz, and although typical surfaces are smooth in the ideal case, noise can change this landscape considerably. 
Fig.~\ref{fig:vqa-search-space} compares the smooth, convex noise-free optimization surface of ideal VQE to local minimums defined as $Tr[\mathcal{H}_\rho] = \Bra{\psi}\mathcal{H}\Ket{\psi}$ for pure states found in the traditional, noisy and error-mitigated search regions. As seen in Fig.~\ref{fig:vqa-search-space}, reducing quantum system noise discovers solutions closer to the ideal. Thus, we are motivated to include error mitigation that is tuned dynamically within VQA routines.

\subsection{Error Mitigation Overview}
\label{back-em}

With the limited number of qubits available in near-term quantum computers, full-fledged quantum error correction is impractical. 
QEC techniques require uniting multiple physical qubits on a device to act as a single logical qubit and special codes to remedy errors are implemented on these multiple physical qubits.  
Near-term quantum computing is very limited in the number of available physical qubits meaning that we look at alternative approaches - specifically error mitigation instead of correction.
These techniques reduce the effect of noise on the execution of a quantum circuit on the device.

Multiple orthogonal forms of error mitigation exist to correct different forms of quantum errors, many of which can be used in conjunction to achieve maximum fidelity.
\emph{Pre-processing} mitigation includes techniques such as noise aware mapping~\cite{murali2019noise} which selects the most appropriate physical machine qubits to map the circuit's logical qubits onto.
\emph{Post-processing} techniques can correct measurement errors~\cite{tannu2019mitigating}, extrapolate for zero noise~\cite{giurgicatiron2020digital,li2017efficient,temme2017error, zne4} and so forth.
The errors mitigated by these techniques are largely influenced by machine / circuit characteristics which are externally observable i.e. mostly independent of the internal workings (e.g. quantum state) of the particular quantum circuit executing on the quantum machine.  
They aim to minimize the impact of systematic noise in a ``one-size fits all manner'' and can be incorporated as is, with minimal understanding of the state of the quantum system.

Other mitigation techniques such as dynamical decoupling~\cite{DDBiercuk_2011,jurcevic2021demonstration,DDKhodjasteh_2007,DDPokharel_2018}, spin-echo correction~\cite{hahn} and gate scheduling~\cite{smith2021error} are more dependent on internal quantum state, as they try to refocus signals in open quantum systems and their effectiveness is amplified if deployed in a manner that is optimized to the compiled quantum application characteristics as well as the quantum device noise profile. 
We focus on these open-loop techniques in this work and optimize them via a variational approach; they are discussed in more detail next.

\begin{figure*}[!t]
\centering
\fbox{
\includegraphics[width=\textwidth,trim={0cm 8.25cm 0cm 8.25cm}]{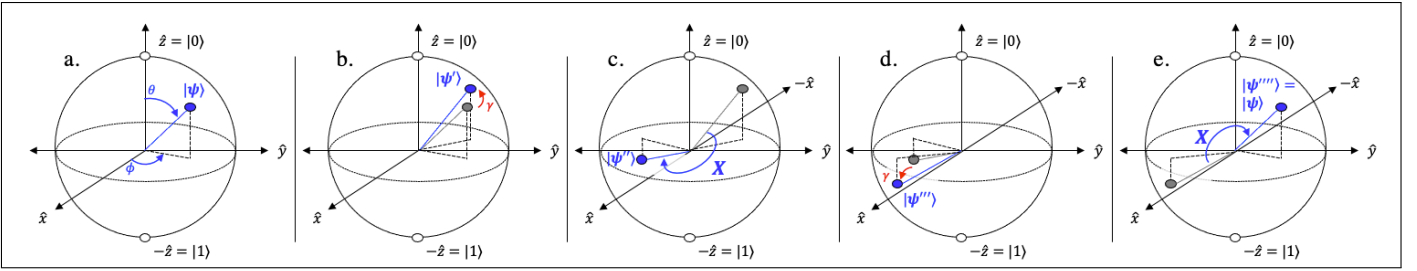}
}
\caption{Phase refocusing through Hahn spin-echo. (a) A qubit state, $\ket{\psi}$, is prepared. (b) As time elapses, the phase of $\ket{\psi}$ decoheres, and noise in the form of $R_z(\gamma)$ creates the quantum state $\ket{\psi'}$. (c) An $\textbf{X}$ gate rotates $\ket{\psi'}$ 180$^\circ$ around the x-axis to create $\ket{\psi''}$. (d) The effects of dephasing, another $R_z(\gamma)$ rotation, constructively interferes with $\ket{\psi''}$ to produce $\ket{\psi'''}$. (e) Another $\textbf{X}$ pulse restores $\ket{\psi}$ from $\ket{\psi'''}$.}
\label{fig:xx_correction}
\end{figure*}

\section{Idling Errors and their Error Mitigation}
\label{idle}

In this work we target two mitigation techniques which are implemented in circuit idle times that appear during runtime.
We define qubit runtime as the period spanning the first gate up until measurement. 
During runtime, a qubit will spend some cycles active within computation and others idle while waiting for signals to propagate along a critical path. 
Limited connectivity in near-term devices requires routing networks for qubit communication in mapped circuits, and these routing networks are inserted during the compilation of machine-agnostic algorithms into machine-ready executables.  
Logically, communication constraints of nearest-neighbor QCs will cause an increase in idle time as algorithms scale. 
As idle time increases, qubit state can decohere, leading to poor circuit fidelity.

The most elementary form of idle time error mitigation suppresses single-qubit phase accumulation with Hahn spin-echo techniques where $R_x(\pi) = X$ instructions are insert into circuits during runtime. These instructions reverse the impact that dephasing has on quantum state over time. For example, consider a quantum state $\ket{\psi}$ pictured on the Bloch sphere in Fig.~\ref{fig:xx_correction}(a) that is prepared with a rotation of $R_x(\theta)$ and $R_z(\phi)$. Ideally, $\ket{\psi}$ would hold state information indefinitely, but phase information is susceptible to decoherence. In Fig.~\ref{fig:xx_correction}(b), the decay of state information is represented by the unknown rotation $R_z(\gamma)$ that causes $\ket{\psi}$ to evolve to $\ket{\psi'}$. Hahn spin-echo techniques apply an $X$ operation to $\ket{\psi'}$ in Fig.~\ref{fig:xx_correction}(c) to mitigate the phase accumulation caused by decoherence, resulting in state $\ket{\psi''}$. Continued dephasing in the form of another $R_z(\gamma)$ is shown in Fig.~\ref{fig:xx_correction}(d). This accumulation of more phase counteracts the original rotation of $R_z(\gamma)$, refocusing the qubit state to create $\ket{\psi'''}$. After the application of a second $X$ gate, the original state of $\ket{\psi}$ is restored (Fig.~\ref{fig:xx_correction}(e)). The procedure of inserting $R_x(\pi)R_x(\pi) = XX$ mid-circuit is permitted as it preserves the semantics of the original circuit. The $X$ operation is self-inverse and 
 $XX^{-1}=I$ where $I$ is the identity operation.

The two idling error mitigation techniques we focus on, both inspired by the Hahn spin-echo techniques from above are Dynamical Decoupling and single-gate scheduling within idle windows.

\subsection{Dynamical Decoupling}
\label{back-dd}
Dynamical decoupling (DD)~\cite{viola1999dynamical}  ``decouples'' compute qubits from environmental noise with special gate sequences that do not change the overall logic of a circuit. The $XX$ sequence implements the most basic form of DD that provides Hahn spin-echo corrections. Standard DD with the ``universal decoupling'' sequence requires four gates: $R_x(\pi)R_y(\pi)R_x(\pi)R_y(\pi) = XYXY$~\cite{viola1999dynamical}. 
The universal decoupling sequence (referred to as $XY4$) adds increased protection to state because $\theta=\pi$ rotations about both the x and y axes makes the qubit more resilient to environmental noise~\cite{souza2012robust}.
Other DD sequences include $YY$, $XY8$, and in general $XX....XX$, $YY....YY$, $XYXY....XYXY$ (to N) wherein identifying the optimal number of DD sequences N within an idle window is an unsolved problem.
More details and other sequences are discussed in prior work~\cite{DDPokharel_2018}.

DD has proven effective at correcting single qubit states and, to a lesser extent, two qubit entangled states in superconducting systems~\cite{pokharel2018demonstration}. 
DD has also improved the Quantum Volume (QV) of a real QC. Both of these demonstrations, however, cost additional circuit instructions during runtime. 
Single-qubit gate errors on superconductors are on average of order $10^{-4}$~\cite{IBMQE,jurcevic2021demonstration}, and although individually small, collective errors can degrade circuit performance, especially as circuits scale on maximally-utilized machines.

Thus, if DD is applied to circuit optimization, it must be included within intelligent routines that avoid introducing additional gate error that outweighs corrective benefits; the optimal number of DD sequences to be inserted in an idle window is largely unknown and VAQEM's variational approach can find this optimal sequence length effectively.

\subsection{Single-qubit Gate Scheduling}
\label{back-gate}
As late as possible (ALAP) scheduling tends to be standard in compilation tools meaning that gates will not execute until another operation, typically either a measurement or a two-qubit operation along a critical path, can be executed immediately afterwards. Scheduling qubit operations for ALAP assists with mitigating noise associated with $T_1$ and $T_2$ decoherence if qubit runtime has not initialized. ALAP execution, however, is not always ideal once a qubit holds state. 
DD techniques use additional gates to recohere quantum state in the presence of noise. Rather than add gates to a circuit, single-qubit gate scheduling refocuses signals by appropriate scheduling of gates within idle windows. 
Prior work on single-qubit gate scheduling~\cite{smith2021error} builds tuning circuits from the original circuit by slicing the circuits into many parts and appending the reversal of the circuit to the slice. It then leverages the reversible nature of quantum computation to find optimal gate positions for single-qubit gates adjacent to slack within a quantum circuit.
While effective in the general case where circuit outcomes are unknown, the process requires the construction of tuning circuits and is limited by maximum circuit depth.
VAQEM's variational approach can find optimal gate positions more accurately, throughout the circuit and with less overhead.

\section{Tuning the error mitigation}
\label{TEM}
In this section, we show how tuning specific features of error mitigation techniques can improve their efficacy over a one-size-fits-all approach.

\begin{figure}[t]
\centering
\fbox{\includegraphics[width=0.95\columnwidth,trim={0cm 0cm 0cm 0cm}]{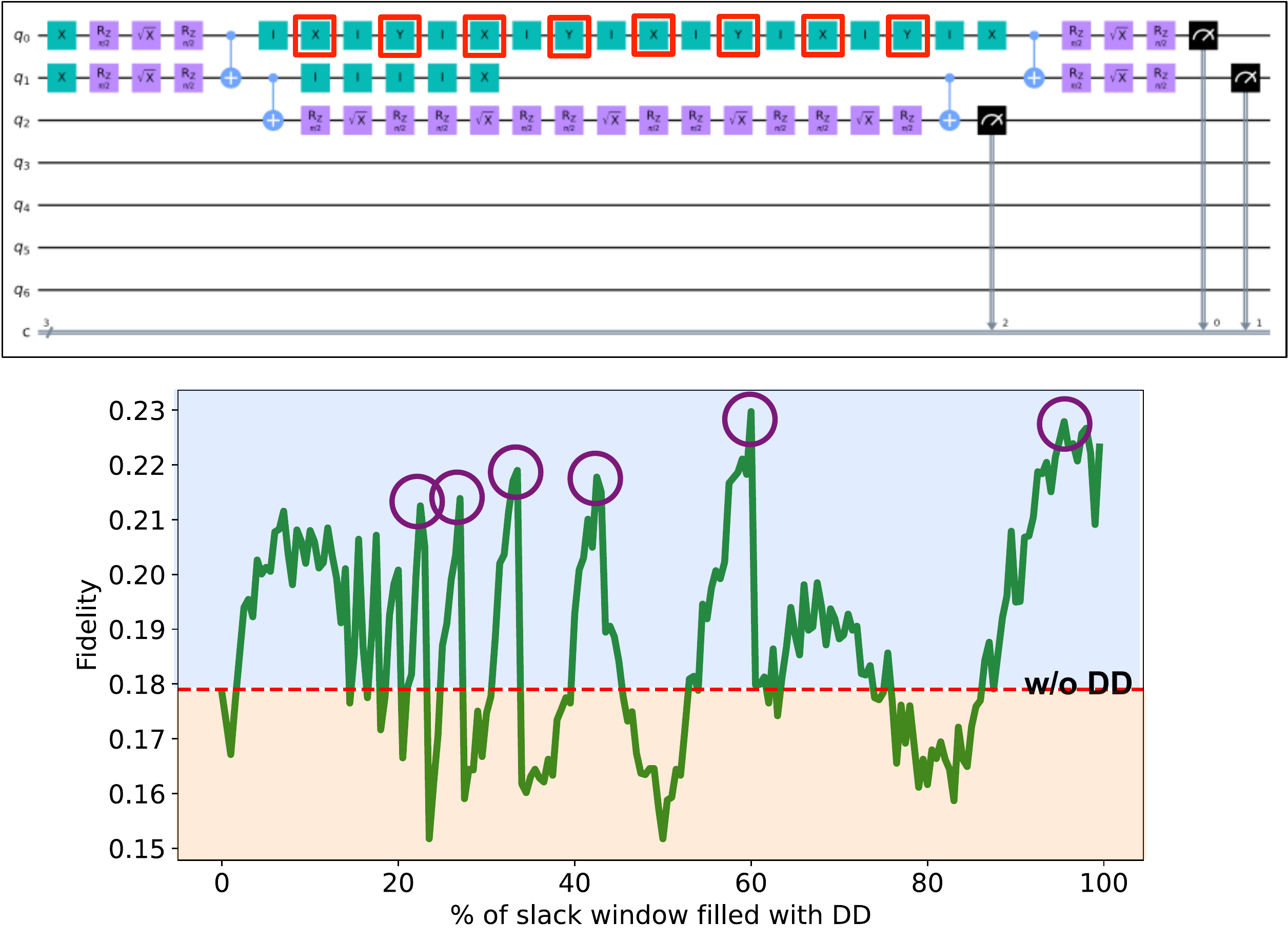}}
\caption{Demonstration of adding DD sequences for error mitigation. The circuit on the top shows the addition of the 2 sets of the XYXY (red boxed) sequence in an idle window. The graph below shows that as the number of such sequences are varied from none to filling the entire idle slack window, they have distinct impact on the circuit's fidelity. The red line indicates fidelity without DD and some sequences show decreased fidelity (yellow region). Blue region implies fidelity gains and peaks that can be found via variational tuning are circled.}
\label{fig:dd-rounds}
\end{figure}

\subsection{Dynamical Decoupling}
First, we explore tuning the number of DD sequences inserted into circuit idle  windows.
As alluded to in Section \ref{back-dd}, the optimal number of DD sequences is dependent on multiple factors such as the input signal, gate errors, impact of crosstalk, qubit quality, and much more. 
Further, it is dependent on the DD sequence itself.
Thus, it is arduous to derive an exact number from theory.
The circuit shown in Fig.~\ref{fig:dd-rounds} shows an example of DD sequence insertion - the insertion of $XYXY$ gates within one large idle window of the circuit.
The graph below it shows that when the number of such $XYXY$ sequences are varied (from none to filling the entire windows), the impact on overall fidelity is distinctly different. 
The red line indicates fidelity without DD and some sequences show decreased fidelity (yellow region). 
Decrease in fidelity can be a byproduct of cumulative gate errors as well as ineffective signal correction.
On the other hand, increase in fidelity (i.e. the blue region) is more prominent.
Some number of sequences are more favorable than others leading to multiple fidelity peaks.
The variational error mitigation tuning which optimizes towards a known target objective function (described in Section \ref{Design}) is able to identify such prominent peaks, thereby optimally improving fidelity with the optimum amount of DD.

\begin{figure}[t]
\centering
\fbox{\includegraphics[width=0.95\columnwidth,trim={0cm 0.2cm 0cm 0.2cm}]{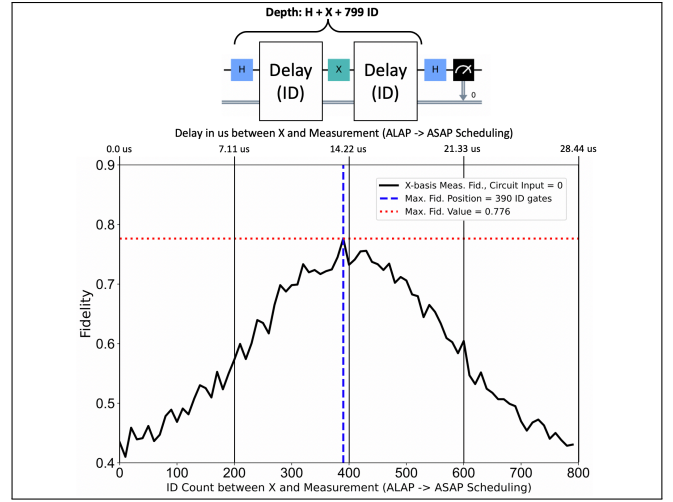}}
\caption{Qubit dephasing correction via Hahn spin-echo techniques. The pictured $H$+$X$+Delay circuit on a single qubit  
tunes X gate placement within a slack window to relate position to state fidelity. An $H$ at the circuit end causes an X-basis measurement ($\Ket{+}/\Ket{-}$), capturing information about qubit phase. When $X$ is scheduled near the middle of the slack window with a 390 $ID$ delay (approx. 14.22$\mu$s), the Hellinger fidelity comparing experimental distributions to the ideal is maximized.}
\label{fig:hahn-echo-test}
\end{figure}

\subsection{Single-qubit Gate Scheduling}

We are inspired by the theory of Hahn spin-echo to reschedule single-qubit gates within idle windows for decoherence mitigation~\cite{smith2021error}.
The circuit in the top of Fig.~\ref{fig:hahn-echo-test} provides a micro-benchmark that demonstrates the viability of optimizing single-qubit gate execution when an adjacent idle window exists. An IBM QC was used for micro-benchmark experiments. 

The circuit begins with an $H$ gate that puts a qubit into superposition of $\Ket{\psi}=\frac{\ket{0}+\ket{1}}{\sqrt{2}}$. Next, an idle window of duration 28.44$\mu s$ artificially created with 799 identity ($ID$) operations (each of approximately $35.56 ns$ duration) and an $X$ gate that is tuned in position within the idle window so that it sweeps each position from ALAP to as soon as possible (ASAP) within the window. As a note, the $X$ gate swaps the amplitude of $\Ket{0}$ and $\Ket{1}$, leaving the overall state of $\Ket{\psi}=\frac{\ket{0}+\ket{1}}{\sqrt{2}}$ unchanged.  
To capture the impact of qubit dephasing, an $H$ at the end of the circuit allows measurement to be in the X-basis ($\Ket{+}=\frac{\ket{0}+\ket{1}}{\sqrt{2}},\Ket{-}=\frac{\ket{0}-\ket{1}}{\sqrt{2}}$) rather than in the Z-basis ($\Ket{0},\Ket{1}$). Measurement in the X-basis provides insight about the dephasing of $\Ket{\psi}$.

The circuit in Fig.~\ref{fig:hahn-echo-test} is inspired by $T_2$ experiments, and it demonstrates that single-qubit gate placement within idle windows impacts fidelity.
We define fidelity as the Hellinger fidelity between an ideal distribution and one produced from a real QC run. The graph in the lower half of Fig.~\ref{fig:hahn-echo-test} demonstrates that gate placement within slack can influence circuit outcome. 
When $X$ is scheduled near the center of the window, fidelity is maximized. 
The micro-benchmark shows the effectiveness of rescheduling within a single window.
Ideally, VAQEM would use the variational approach to tune gate schedules across all idle windows in unison to find a synergistic combination of parameters.

\section{Soundness of Tuning Error Mitigation}



As seen in Fig.~\ref{fig:vqa-search-space}, the impact of environmental noise causes VQA convergence on local energy minimums that are higher than the global optimum. 
Thus, reducing noise is key. 
Error mitigation that reduces external environmental influence on computation is often treated as a separate processing task run sequentially or in parallel to VQA parameter tuning.
In this work, we propose including error mitigation \emph{within} the variational framework of VQA.

It is critical that the variational error mitigation tuning does not create any artificial effect which causes the algorithm to return better than ideal solutions. This is ensured by the following:
\circled{1}\ This approach is only suited to purely quantum mitigation strategies that act on quantum qubits / gates / noise channels. 
\circled{2}\ Per theoretical design, one or more of the pure quantum states that can be achieved by the VQA ansatz can, at best, correspond to the ideal ground state, i.e.  the solution to the VQA problem.
\circled{3}\ A noisy quantum state is a sum of probabilities across multiple pure states. So a quantum error mitigation strategy can, at best, reduce the noisy state to the pure state(s) that correspond to the ideal ground state, but never ``better" than the ideal ground state. Thus, VAQEM can, at best, achieve the ideal solution.

Now we prove the soundness of this variational error mitigation approach to not exceed the true global optima using VQE as an example, i.e. in Fig.\ref{fig:vqa-search-space} the green boxes never fall below the blue line.

\textbf{Setup}. Let $\mathcal{H}$ be a Hamiltonian with eigenvectors $\ket{\psi_i}$ and eigenenergies $\textbf{E}_i$ so that $\mathcal{H} \ket{\psi_i} = \textbf{E}_i \ket{\psi_i}$. Let $\textbf{E}_0$ denote the ground state energy with corresponding ground state $\ket{\psi_0}$.  $E_0$ is found on the blue curve representing the ideal optimization surface in Fig.~\ref{fig:vqa-search-space}. For simplicity, assume that the ground state is degenerate, meaning no other state has energy of $\textbf{E}_0$ (the analysis still holds if we relax this assumption).

\begin{theorem} \textbf{(Pure State VQE)}. $\bra{\phi} \mathcal{H} \ket{\phi} \geq \textbf{E}_0$ for every state $\ket{\phi}$, with equality only achieved for the ground state $\ket{\phi} = \ket{\psi_0}$. This is the core motivation behind VQE: one can only overestimate the ground state energy. With a good ansatz, one hopes to approach $\textbf{E}_0$.
\end{theorem}

Proof: Re-statement of the \textit{variational principle}  \cite{shankar2012principles}.

\begin{theorem}
 \textbf{(Mixed State VQE)} The energy of a mixed state is at least $\textbf{E}_0$, so one cannot ``cheat'' by tuning non-unitary operations. A result of VQE for a specific $\mathcal{H}$ must not fall into the region beyond the blue lower bound pictured in Fig.~\ref{fig:vqa-search-space}. More formally, $\text{Tr}[\mathcal{H}\rho] \geq \textbf{E}_0$.
\end{theorem}

Proof: the energy of a mixed state $\rho$ is $\text{Tr}[H\rho]$. By the spectral theorem, $\rho = \sum p_i \ket{\phi_i} \bra{\phi_i}$, where $p_i$ is a probability vector and $\ket{\phi_i}$ are pure states. Therefore,
$$ \text{Tr}[\mathcal{H}\rho] = \sum_i p_i \text{Tr}[\mathcal{H} \ket{\phi_i}\bra{\phi_i}] = \sum_i p_i \bra{\phi_i} \mathcal{H} \ket{\phi_i}$$

\noindent by cyclicity of trace. By Theorem 1, each $\bra{\phi_i} \mathcal{H} \ket{\phi_i}$ term is at least $\textbf{E}_0$. Since $p_i$ is a probability vector, the energy $\text{Tr}[\mathcal{H}\rho]$ is at least $\textbf{E}_0$. Equality is achieved in the special case where the mixed state $\rho$ is actually the pure ground state $\ket{\psi_0}\bra{\psi_0}$.

\section{Designing VAQEM for today's quantum cloud}
\label{Design}

In Section \ref{TEM} we showed that tuning specific features of error mitigation techniques can significantly increase their impact on quantum circuit execution resulting in improved fidelity.
Fidelity increase leads to improvements in the measured objective function for VQA applications.
In this section we show how these error mitigation features are tuned within the VQA harness - we discuss how such features can be optimally tuned in the future (as quantum service offerings improve) and the challenges pertaining to today's machines and corresponding design choices that make VAQEM feasible today even if sub-optimal.

\begin{figure}[t]
\centering
\fbox{\includegraphics[width=\columnwidth,trim={0cm 0cm 0cm 0cm},clip]{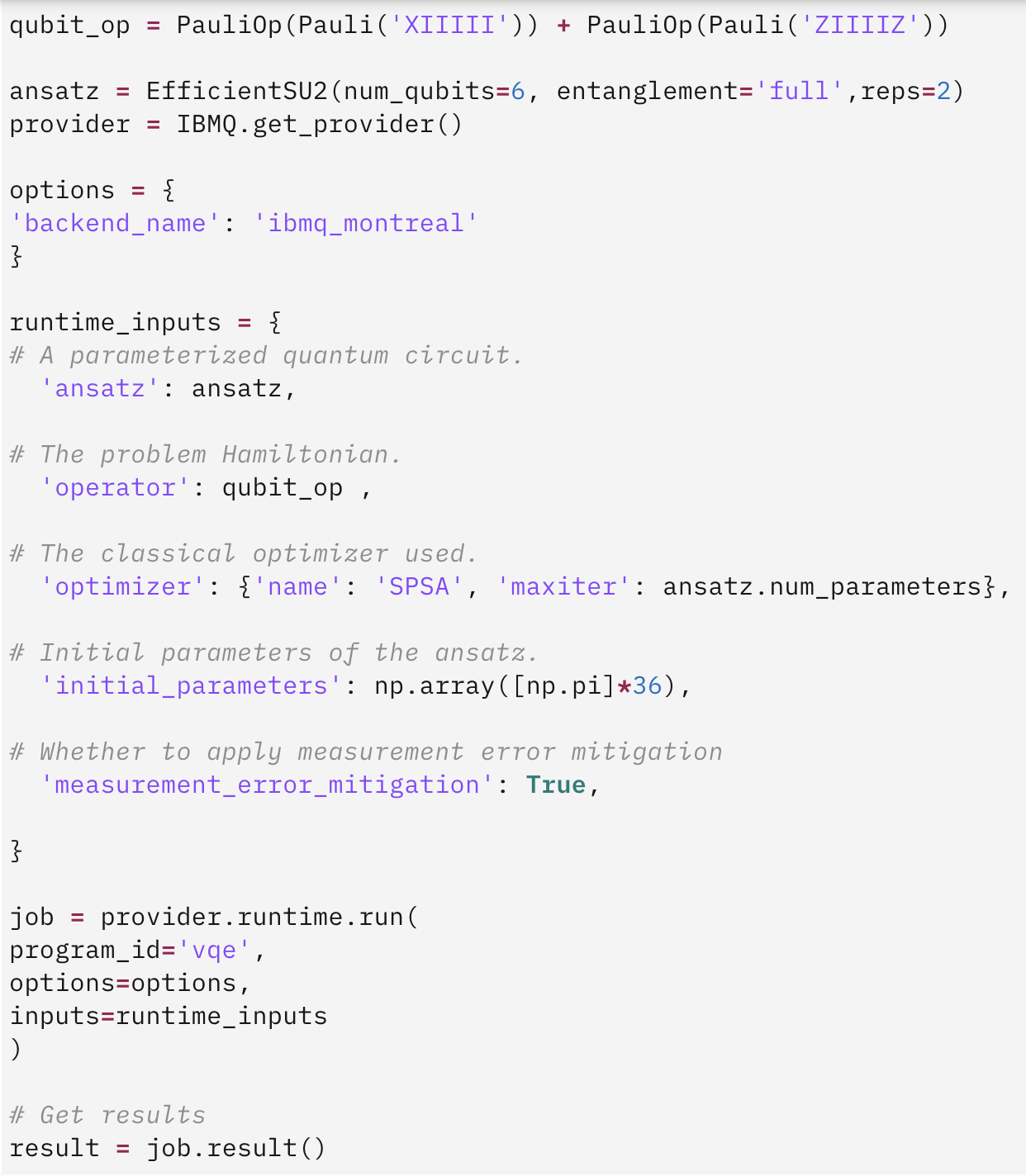}}
\caption{A snapshot of Qiskit Runtime (07/2021) implementing a VQA workload.}
\label{fig:Runtime}
\end{figure}

\subsection{Qiskit Runtime and its Constraints}
In the past, quantum-classical co-processing experiments (like VQA) on IBM machines and beyond, had to either be run entirely via simulation or required (thousands or even millions of)  costly interactions between the client's classical processor and the quantum machine. 
These constant interactions were required to transfer the quantum measurements to the classical computer, which were used to calculate the objective function and then enabling tuning ansatz parameters for the next iteration of the VQA process.
With such a setup, the process of modeling LiH would take as many as 45 days!~\cite{IBMQR}.

IBM Qiskit Runtime~\cite{IBMQR} recently demonstrated a 120x speedup in simulating molecules primarily thanks to the ability to run quantum programs entirely on the cloud. Improving this speed is critical for large calculations and is an active area of work~\cite{clops}.
To the best of our knowledge, this work is among the earliest to utilize the IBM Qiskit Runtime platform in a research endeavor. 
A snapshot of the environment / code sample is shown in Fig.\ref{fig:Runtime}.
The general methodology to implement this variational approach was already shown in Fig.~\ref{fig:vqa}.

New platforms come with constraints though.
Qiskit Runtime is limited\footnote{Limitations correspond to 07/2021 when we were among the earliest users of Qiskit Runtime. Please refer to \url{https://qiskit.org/documentation/partners/qiskit_runtime/tutorials.html} for information on recent updates.} in terms of: \circled{1}\ which parameters can be tuned variationally - only the traditional gate angle parameters are allowed, \circled{2}\ which classical tuners can be used - only the Simultaneous Perturbation Stochastic Approximation method (SPSA)~\cite{SPSA}  variants are allowed, which are usually slow to converge as they only take incremental local steps from start to finish, \circled{3}\ the maximum runtime is limited - we can at most run a problem for only 5 hours which is especially detrimental when using a slow tuning algorithm and larger circuits, and \circled{4}\ machine access is very limited - only one machine with Qiskit Runtime enabled is accessible to us.

While we obtain promising results with the use of Qiskit Runtime (smaller chemistry problems), considering the challenges outlined above, we are forced to pursue alternative methodologies for some steps / scenarios which are unsuited to Runtime (at the present): circuits which require tuning times of more than 5 hours, when machine access is limited, and for tuning the mitigation approaches. 
These are discussed next.

\subsection{Simulation Feasibility}
Simulation is insufficient for scalable quantum computation.
In fact, that is the primary motivation for QCs.
But for the size of the VQA problems which are currently explored (due to machine limitations), simulation is feasible \emph{as long as it accurately models the effects of the quantum machine.}
In this section, we show that while simulation can be appropriate for variationally tuning the VQA angle parameters, they are unsuitable for tuning error mitigation.

\begin{figure}[t]
\centering
\fbox{\includegraphics[width=0.95\columnwidth,trim={0cm 0cm 0cm 0cm},clip]{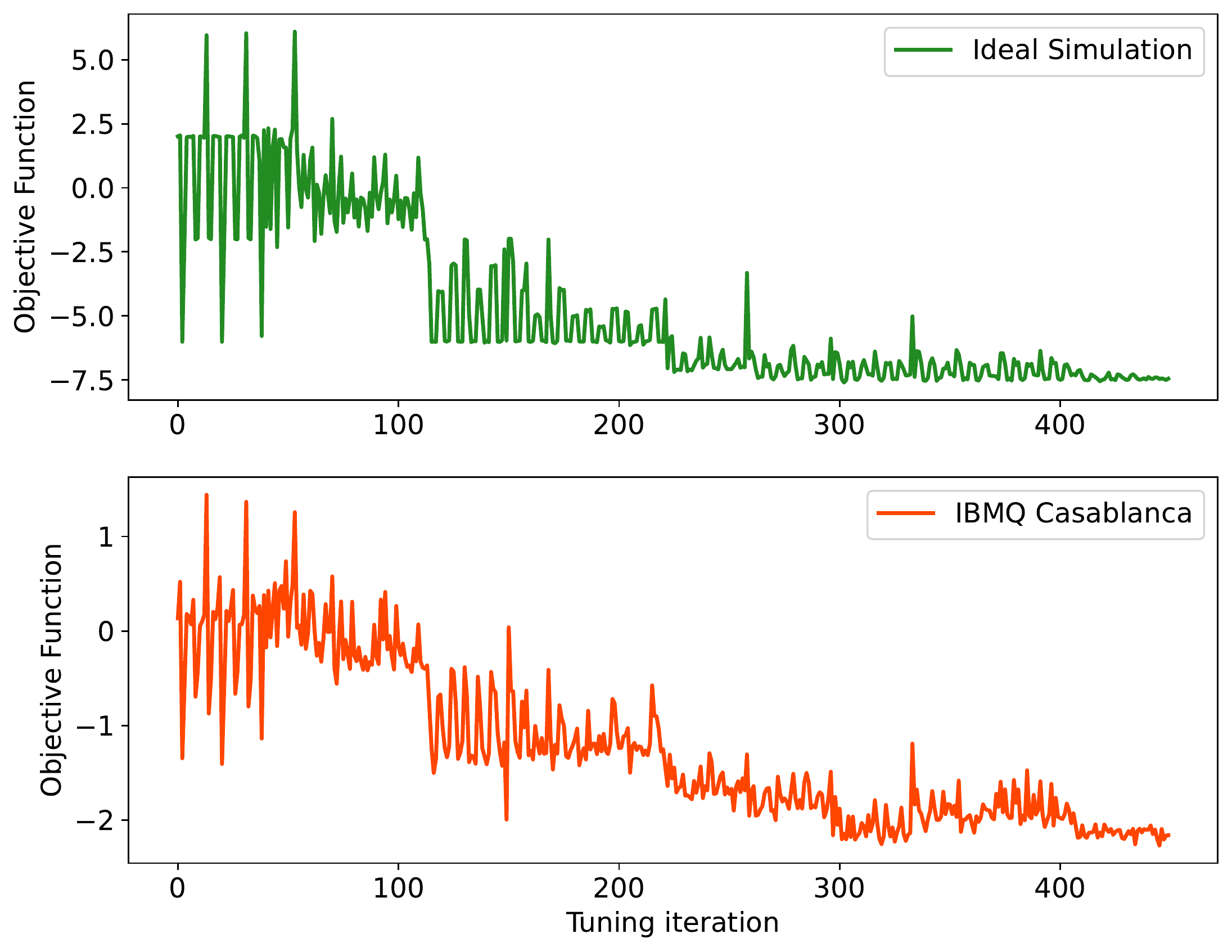}}
\caption{Ideal Simulation and real quantum machine execution for tuning gate rotational angles for a 6-qubit VQE problem. Similar convergence trends are seen on both, even though the objective function value measured varies. Finding optimal minima parameters through simulation are also reasonable minimas for the real quantum machine.}
\label{fig:SimRM-Angles}
\end{figure}

\begin{figure}[t]
\centering
\fbox{\includegraphics[width=0.95\columnwidth,trim={0cm 0cm 0cm 0cm},clip]{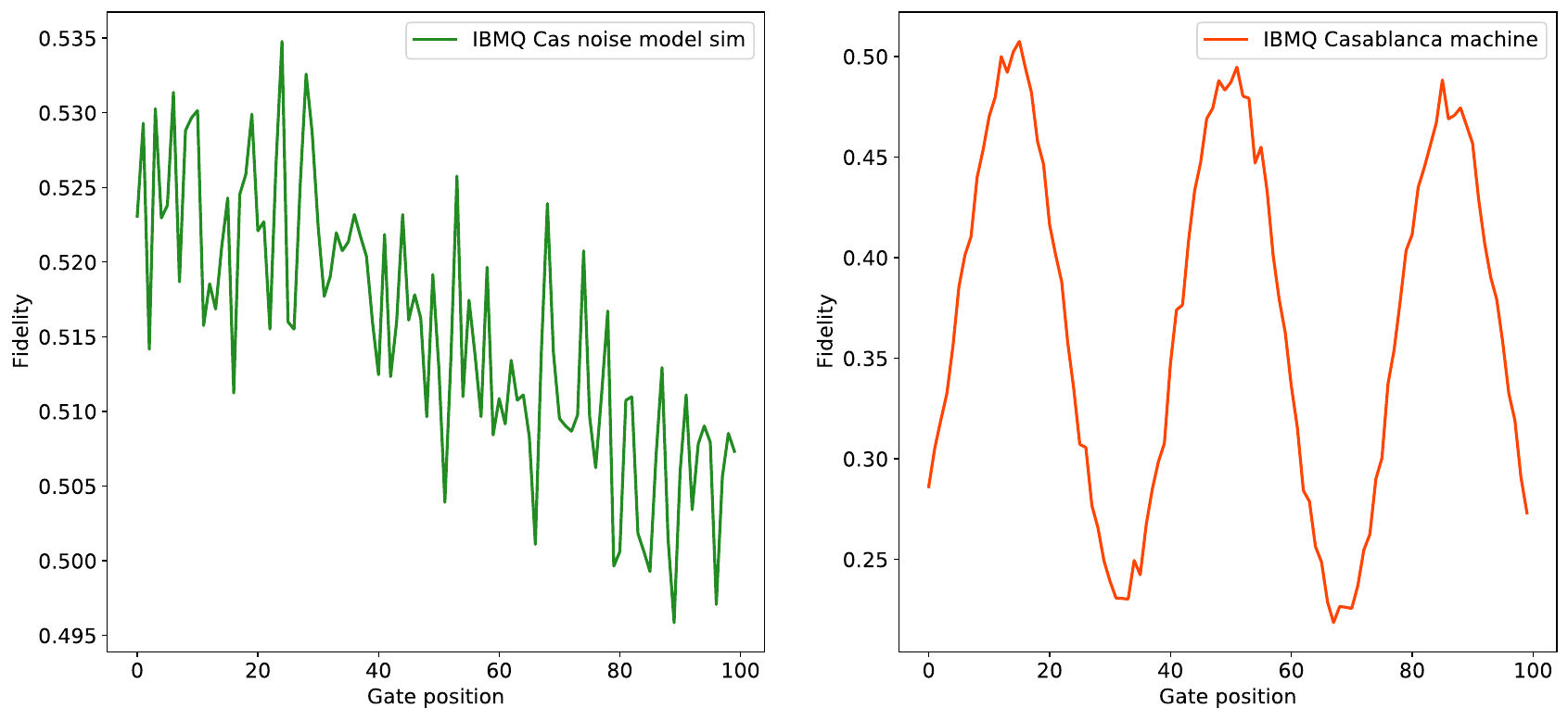}}
\caption{Noisy Simulation and real quantum machine execution for error mitigation via tuning gate positions in idle windows for a 2 qubit microbenchmark with one large idle window. Clearly the trends and range seen in simulation are vastly different from the real machine.}
\label{fig:SimRM-EM}
\end{figure}

\subsubsection{Tuning Angles}
Fig.~\ref{fig:SimRM-Angles} shows the variational tuning of the traditional gate angle parameters for a 6-qubit VQE problem. The top graph shows tuning via ideal simulation while the bottom graph shows results from real quantum machine execution (on {\tt ibmq\_casablanca}). 
Importantly, note that the angle tuning is performed on the ideal simulator and the same tuning parameters are run on the quantum machine. 
It is evident from the graphs that even though the range of the objective function values, as well as the final converged objective measurement are different, the tuning and convergence trends are similar on both. This empirically shows that finding optimal minima parameters through simulation are also reasonable minimas for the real quantum machine.
This Optimal Parameter Resilience in the face of incoherent noise, such as decoherence processes and readout errors has also been motivated from a theoretical perspective in prior work~\cite{Sharma_2020}. 
Thus, simulation is suitable for angle tuning when machines are unavailable / unsuitable.

\subsubsection{Tuning Error Mitigation}
On the other hand, simulation can be less suitable to many forms of error mitigation.
This is because the noise models used in today's simulations are unable to capture a lot of the noise characteristics and interactions within the real quantum machines.
Fig.~\ref{fig:SimRM-EM} shows machine-based noisy simulation ({\tt ibmq\_casablanca} model) and real quantum machine execution (on {\tt ibmq\_casablanca}) for error mitigation via tuning gate positions in idle windows for a 2-qubit micro-benchmark with one large idle window. 
Note that the noise model is obtained from the same calibration cycle as the real device.
Clearly the trends seen in simulation are vastly different from the real machine. While the simulation shows a preference for gate positioning at the beginning of the window, the real machine shows poor fidelity at the start but different optimal positions across the window. Further, the range of fidelity is considerably greater on the real machine.
Similar trends can be seen across other mitigation techniques such as DD as well.
Clearly, benefits of mitigation techniques (in general) and tuning mitigation (in specific) should be evaluated on the real machine until we have a better understanding of the machines to build more accurate noise models.

\begin{figure}[h]
\centering
\fbox{\includegraphics[width=0.95\columnwidth,trim={0cm 0cm 0cm 0cm},clip]{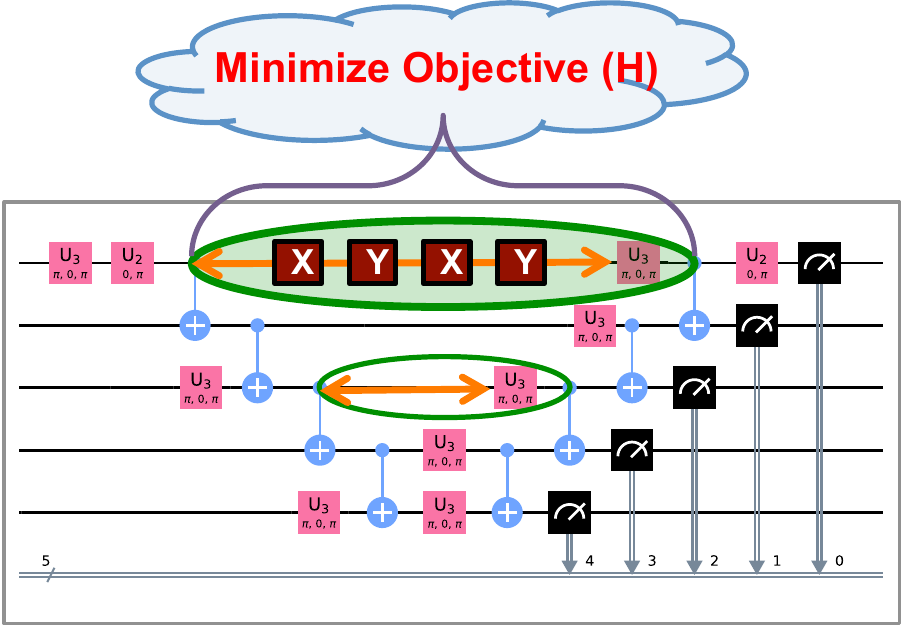}}
\caption{Tuning DD error mitigation on a real QC to discover optimum sequencing that boosts circuit fidelity.}
\label{fig:DD-GT}
\end{figure}

\subsection{Tuning EM on the Machine}

As motivated above, we would like to use the quantum machine where possible - especially when error mitigation tuning is required.
On the one hand, error mitigation tuning cannot be directly performed within Qiskit Runtime in its current form (as discussed earlier).
On the other hand, a distributed feedback based approach, alternating between the client's classical processor and the quantum machine on the cloud is infeasible due to the communication costs.

Thus we take an approach which:  \circled{a}\ first finds the optimum gate rotation angle parameters with methods discussed earlier and then \circled{b}\ tunes error mitigation features in each idle window independently to optimize for the VQA problem's objective function and combines together the optimum tuned values for each window.
Tuning each window independently (i.e. tuning in one window while keeping parameters corresponding to other windows fixed) is a reasonable approach because the error mitigation techniques explored here only involve adding / moving single qubit gates in the idle windows and it is known that impact of single-qubit gate crosstalk is minimal compared to other noise forms.
This is confirmed by our own micro-experiments as well as in prior work~\cite{murali2020software}.

An illustration of the framework performing DD sequence tuning is shown in Fig.~\ref{fig:DD-GT}.
Two windows for EM tuning are circled in green. 
The figure shows a specific window being tuned, shaded with green and DD sequences inserted within the window. The number of DD sequences inserted is swept from none to maximum (i.e. to fill the window entirely) and the objective function is measured for the tuned ansatz.
When this particular window is tuned, other windows are without any DD insertions (i.e. baseline).
The tuning with the lowest objective function value (in a minimization problem) is selected. 
Similarly other windows are tuned and all the optimal tunings are then combined together.
The above technique is also used in the same manner for the single-qubit gate position tuning. 
Note that the resolution of the sweep is constrained by the available resources in the quantum execution framework (discussed further in Section \ref{6-method}).

\begin{figure}[t]
\centering
\fbox{\includegraphics[width=0.98\columnwidth,trim={0cm 0cm 0cm 0cm},clip]{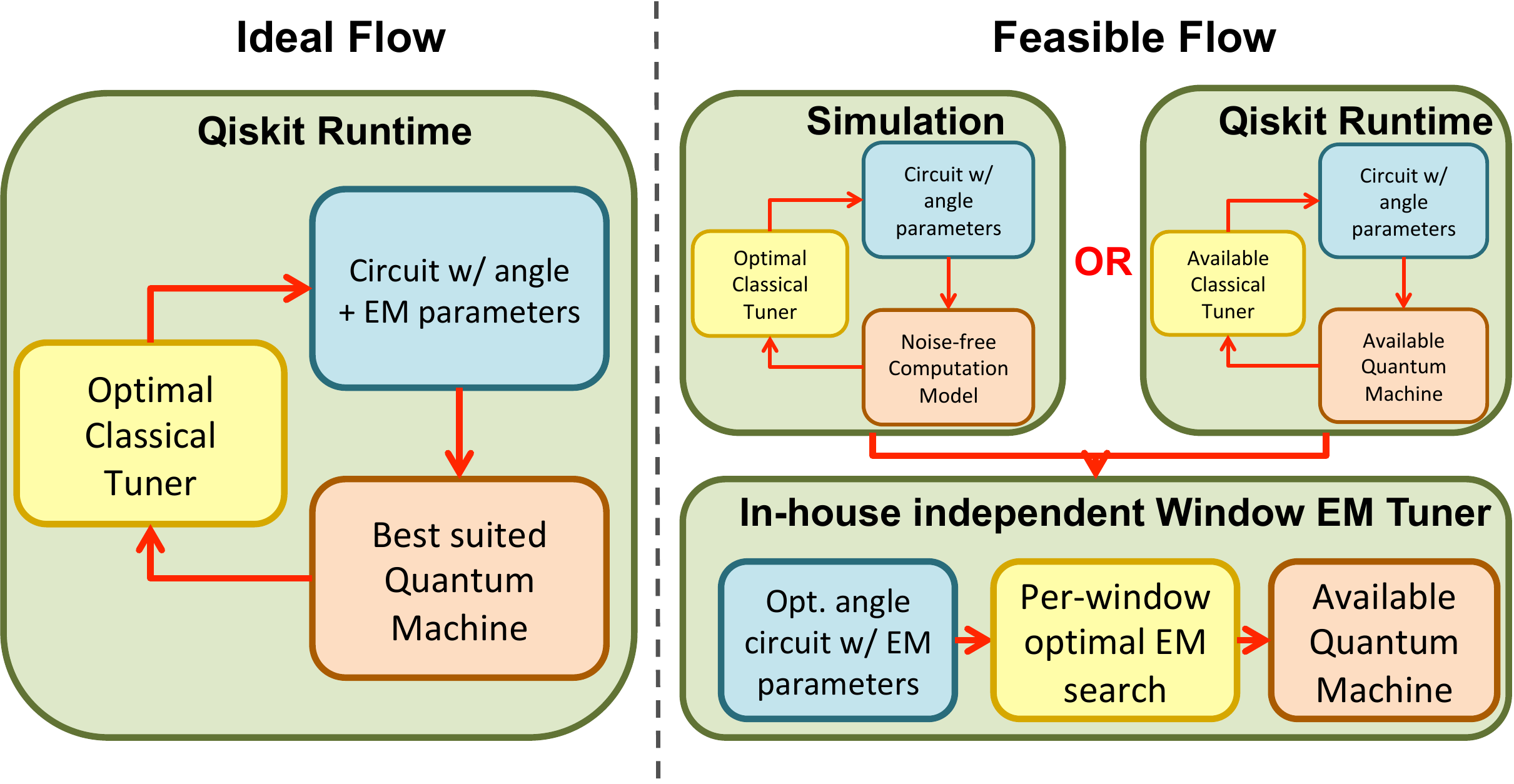}}
\caption{Overall VAQEM flow: Ideal (long term) vs Feasible (near term) methodology.}
\label{fig:flow}
\end{figure}

\subsection{Summary: Overall Workflow}

Fig.\ref{fig:flow} illustrates the overall workflow of VAQEM.
On the left, it shows the ideal flow which we envision utilizing in the future as Qiskit Runtime develops further and supports tuning non-standard variational parameters, allows for more classical tuners and is enabled on more quantum machines.
In such a scenario the entire VAQEM approach can be performed within Qiskit Runtime.

On the right, the figure shows the feasible flow suited to the current IBM quantum cloud to evaluate VAQEM's benefits. 
Tuning the variational gate angles is performed either via simulation or through Qiskit Runtime based on problem tuning feasibility / machine availability. 
In our evaluated applications, we are able to use Qiskit Runtime for two small molecular chemistry applications while the other use simulation for angle tuning (Section \ref{6-method}).
Once the angles are tuned, error mitigation tuning is performed on the real machine again to optimize for the VQA objective function, but using an independent window EM tuner (instead of the traditional feedback based VQA tuning approach).

It is critical to note that simulation as a part of the workflow is only pursued here as a means to proceed to the proposed error mitigation tuning on the real quantum machine - i.e. to showcase the benefits from VAQEM. Simulation is not a scalable solution and is very limited in its capabilities, primarily only suited to today's trivial quantum problems. In fact, if ideal simulation was always possible in non-exponential time, there would be no need to perform error mitigation, or even use a quantum computer at all!


\begin{figure*}[t]
\centering
\includegraphics[width=\textwidth,trim={0cm 0cm 0cm 0cm}]{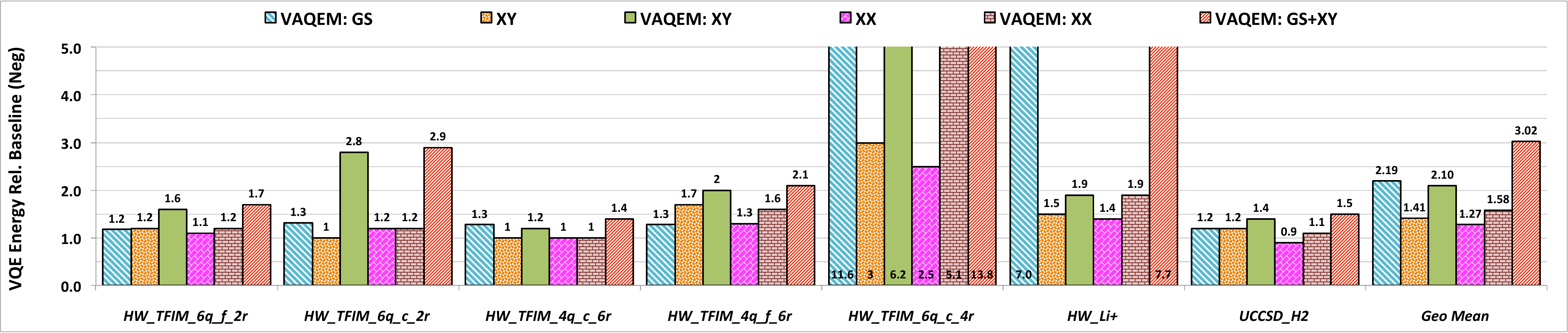}
\caption{Improvements to VQE Energy relative to a baseline with only measurement error mitigation. Higher is better. VAQEM benefits with GS, DD and GS+DD are shown. VAQEM provides benefits over ``naive" error mitigation and, further,  techniques can combine in a synergistic manner for greater improvements.  }

\label{fig:result-1}
\end{figure*}

\section{Methodology}
\label{6-method}


\subsection{Applications}

We limit ourselves to one VQA domain, the VQE which was introduced in Section \ref{VQA}.
Due to machine limitations on circuit width and depth applications are restricted to 6 qubits or fewer and of shorter duration. 
We evaluate 7 different VQA applications built for 3 Hamiltonians:

\circled{1}\ \emph{Transverse Field Ising Model / Hardware Efficient Ansatz:} The one dimensional TFIM is a ubiquitous model which has applications to understanding phase transitions in magnetic materials~\cite{uvarov2020machine}. The TFIM is a desirable system since it is exactly solvable via classical means. We solve for TFIM ground state energy on 5 different hardware
efficient SU2 ansatz~\cite{IBM-SU2}. The ansatz are constructed for 4 qubits and 6 qubits and for ``full'' and ``circular'' entanglement. Further, the number of block repetitions within the ansatz structure are varied between 2, 4 and 6.  

\circled{2}\ \emph{Hydrogen Molecule / UCCSD Ansatz:} Unitary Coupled Cluster Single-Double (UCCSD) is an ansatz motivated by principles of quantum chemistry. Hydrogen, being the smallest molecule, is most suited for UCCSD today due to the higher depth of UCCSD. The UCCSD ansatz used is the Qiskit implementation with the Hartree-Fock initial state, with ``parity'' qubit mapping and without two qubit reduction. This generates a 4 qubit ansatz with 15 Hamiltonian terms, 4 of which were truncated with very negligible coefficients. 

\circled{3}\ \emph{Li+ Ion / Hardware Efficient Ansatz:} The Lithium ion was unsuited to the UCCSD ansatz due to circuit depth. So we instead use the SU2 ansatz. It requires a 6 qubit ansatz, which we use with 3 repetitions and `full' entanglement. This generated 55 Hamiltonian terms, around 25 of which were truncated with very negligible coefficients.

Benchmark circuit depth and number of idle windows targeted by error mitigation are shown in Table \ref{Table1}.
Note that benchmarks with higher depth and more (6) qubits stress the 7q IBMQ machine more - greater decoherence and  accumulation of significant CX errors.

\begin{table}[t]
\resizebox{\columnwidth}{!}{%
\begin{tabular}{|l|l|l|l|l|l|l|l|}
\hline
\textbf{Bench} &
  \textbf{6q/f/2r} &
  \textbf{6q/c/2r} &
  \textbf{4q/c/6r} &
  \textbf{4q/f/6r} &
  \textbf{6q/c/4r} &
  \textbf{Li+} &
  \textbf{H2} \\ \hline
\textbf{Depth} &
  54 &
  31 &
  57 &
  101 &
  55 &
  90 &
  61 \\ \hline
\textbf{\# Win} &
  42 &
  24 &
  22 &
  34 &
  30 &
  45 &
  26 \\ \hline
\end{tabular}%
}
\caption{Benchmark characteristics. The first 5 columns show  TFIM while the last two are from molecular chemistry. ``Depth" is circuit depth in terms of CX gates and ``\# Win" is number of idle windows targeted.}
\label{Table1}
\end{table}

We integrate VAQEM as part of the Qiskit~\cite{Qiskit} framework - it is added as a final step in the compilation process. 
Applications using Qiskit Runtime (the two molecular chemistry applications) use {\tt $ibmq_montreal$} (27 qubits).
The other 5 applications that don't use Qiskit Runtime are distributed across 3 quantum devices: {\tt ibmq\_guadalupe} (16 qubits), {\tt ibmq\_jakarta} (7 qubits), and {\tt ibmq\_casablanca} (7 qubits).
Machine details can be found on the IBM Quantum Systems page~\cite{IBMQS}.

\subsection{Evaluation Comparisons}

\emph{Baseline / MEM:} The baseline uses no DD sequences within the idle windows and schedules single-qubit gates As Late As Possible (ALAP). This is the standard compilation via Qiskit. We add measurement error mitigation (MEM) to improve baseline fidelity. MEM can be applied orthogonal to VAQEM mitigation techniques.

\emph{No-EM:} Similar to the baseline above but without MEM, thus the worst performer among our evaluations.

\emph{DD (XY / XX):} The basic DD designs incorporate a single round / sequence of DD within the idle windows. There are 3 versions, employing XY4, XX respectively. The DD sequence is spread out in the idle window, evenly spaced out i.e. a periodic DD distribution~\cite{DDBiercuk_2011}.

\emph{VAQEM DD (XY / XX):} Proposed variational approach to DD error mitigation by tuning the number of DD sequences inserted in the idle windows based on the VQA objective. Inserted sequences are spaced out as a periodic distribution. 

\emph{VAQEM GS:} Proposed variational approach to single-qubit gate scheduling  by tuning the gate positions in idle windows based on the VQA objective.

\emph{VAQEM GS+XY} Incorporates both DD and GS within the VAQEM framework and both are tuned in a coordinated manner. For DD, we only target XY since it is the best performer among the different sequences.

We evaluate VAQEM based on energy estimates using the standard Hartree Energy metric.

\section{Evaluation Results}

\subsection{Ground State Energy Improvements with VAQEM}

Fig.\ref{fig:result-1} shows the benefits from variationally tuning error mitigation via VAQEM. 
The figure shows improvements to the ground state energy measurements relative to a baseline with only measurement error mitigation which is applied orthogonal to VAQEM.
Note that while the actual energy values are negative, we are showing relative improvements here which are positive and thus, higher is better.
The VAQEM variational approach to tuning gate positions enables a 2.19x better VQE energy estimate on average compared to the baseline.
Improvements as high as 7-11x are seen for two applications but note that the absolute energy numbers can be small so relative improvements can seem magnified.
Energy numbers (and circuit fidelity) are especially low for \emph{HW\_TFIM\_6q\_c\_4r} because the circuit is deep and it is forced to use all but one qubit on the device (i.e. hence using noisy qubits as well). Details in Section \ref{6-method}.
Deeper circuits also provide more (and longer) idle windows, increasing the potential for optimal error mitigation benefits.

Two basic DD techniques are also shown, which were discussed in Section \ref{6-method}.
They are able to achieve 27-41\% improvements over the baseline.
XY4 sequences have better benefits as was discussed earlier in Section \ref{back-dd}.
Applying the VAQEM variational approach to the DD techniques increase their benefits by 31-69\%, improving the VQE energy measurements by 1.58x-2.1x (rel. baseline).
As discussed above, benefits are higher for deeper circuits.

The last bar shows the scenario which combines both gate scheduling and dynamic decoupling within the VAQEM framework.
VAQEM enables the techniques to only interact constructively - the tuner will weed out any destructive interference.
In combination, the approach achieves 3.02x energy improvements over the baseline - considerable improvements like this are critical to advance variational algorithm usecases.

\begin{figure}[t]
\centering
\includegraphics[width=0.98\columnwidth,trim={0cm 0cm 0cm 0cm}]{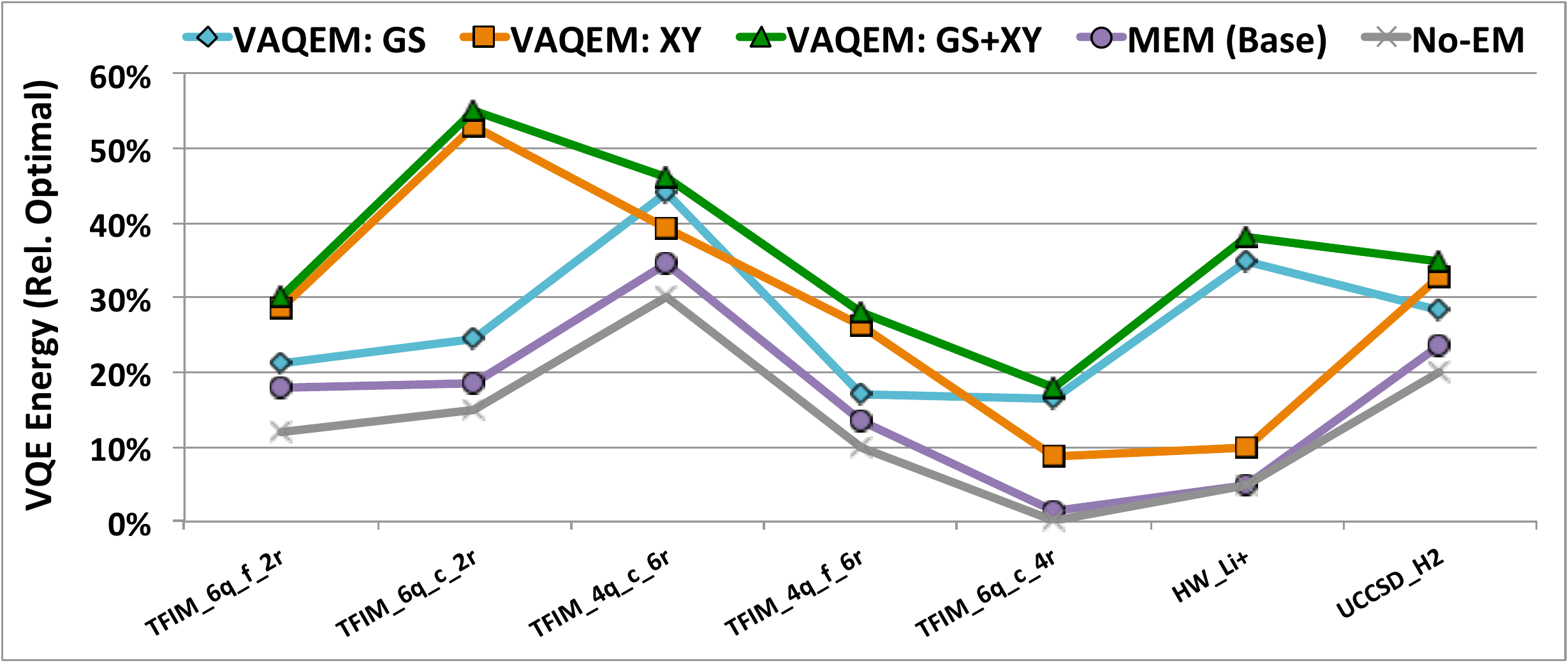}
\caption{VQE energy measurements relative to the optimal value obtained from simulation. Results are shown for single and combined mitigation via VAQEM, baseline with measurement mitigation, and energy under no mitigation.}

\label{fig:result-2}
\end{figure}

\subsection{Energy Measurements Relative to Simulated Optimal}

Next, in Fig.\ref{fig:result-2} we show the VQE energy measurements relative to the optimal value obtained from ideal classical simulation.
We show results for different VAQEM strategies, for the baseline with MEM and for the minimum no-EM scenario.

Measured energy can be very dependent on noise and machine characteristics.
Note that the current target applications are small enough that they can be simulated classically for comparison - but this will not scale to larger circuits and is the fundamental motivation for quantum computers.

The No-EM approach achieves ground state energy estimates of only 1-30\% of optimal while the MEM baseline improves to 2-35\%.
The VAQEM DD approach with XY gates (VAQEM: XY) is able to achieve 10-52\% of the optimal while VAQEM gate scheduling approach (VAQEM: GS) achieves 17-45\% of the optimal.
Finally, the combined VAQEM approach incorporating both DD and GS (VAQEM: GS+XY) always performs best, achieving 19-55\% of the optimal.

While the benefits from the variational approach to the two chosen mitigation techniques are evident, clearly there is room for significant improvement beyond these.
Other mitigation techniques can be employed as part of the variational framework, as well as orthogonally, outside of it.
We reiterate that incorporating techniques into the variational framework always ensures synergistic improvements because any negative interactions are avoided by the tuning mechanism.


\begin{figure}[t]
\centering
\includegraphics[width=0.98\columnwidth,trim={0cm 0cm 0cm 0cm}]{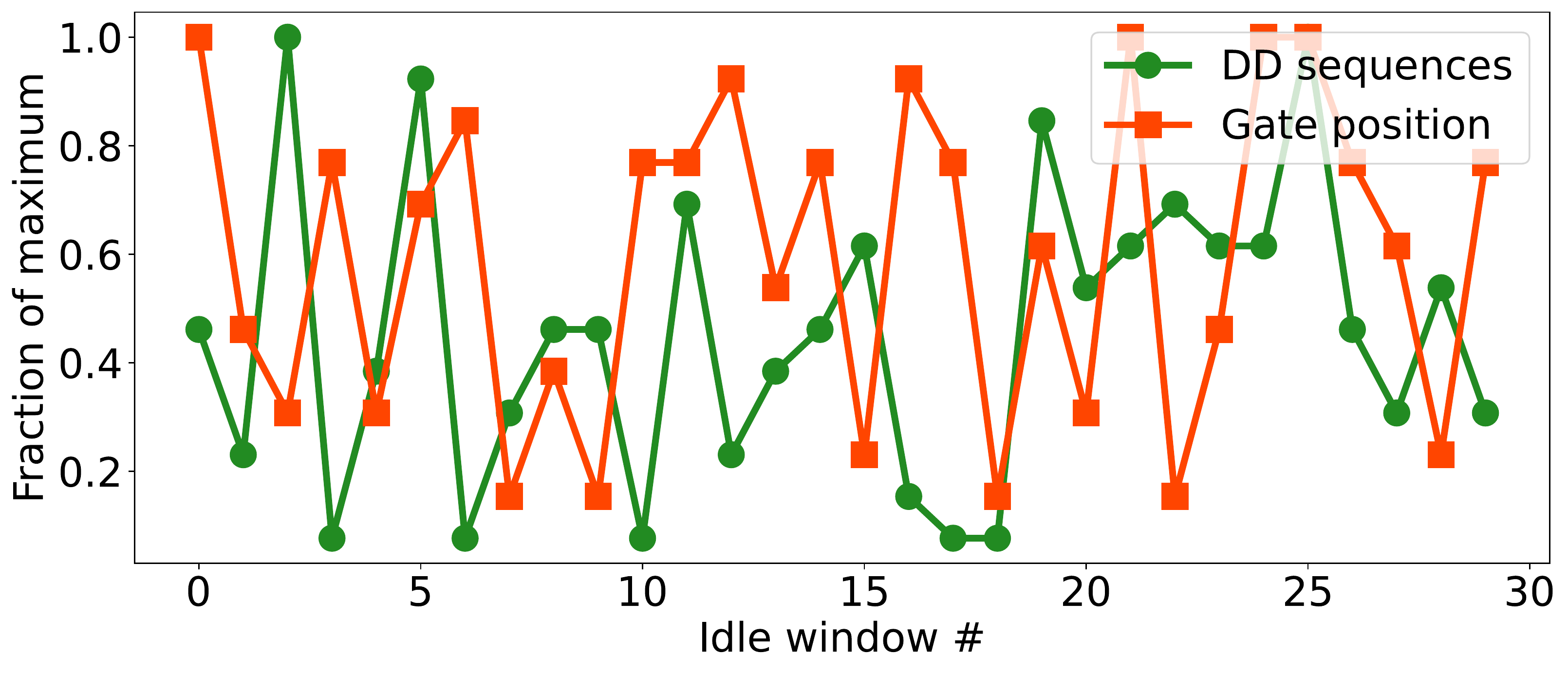}
\caption{Gate positions and number DD sequences (as fractions of maximum) in idle windows of HW\_TFIM\_6q\_c\_4r.}
\label{fig:result-x}
\end{figure}

\subsection{Analyzing VAQEM circuit impact}
Fig. \ref{fig:result-x} shows the gate positions as well as the number of DD sequences (as a fraction of maximum possible), within each idle window of HW\_TFIM\_6q\_c\_4r, as chosen by VAQEM.
It can be observed that the chosen gates positions and the number of DD sequences widely vary across the idle windows.
This is important to highlight the usefulness of the variational approach. 
Each idle window has a different optimal error mitigation ``configuration" which is dependent on input state to the idle window, the qubit characteristics etc.
And it is not trivial to identify the optimal configuration, clearly motivating the variational approach that VAQEM proposes.

\begin{figure}[t]
\centering
\includegraphics[width=0.98\columnwidth,trim={0cm 0cm 0cm 0cm}]{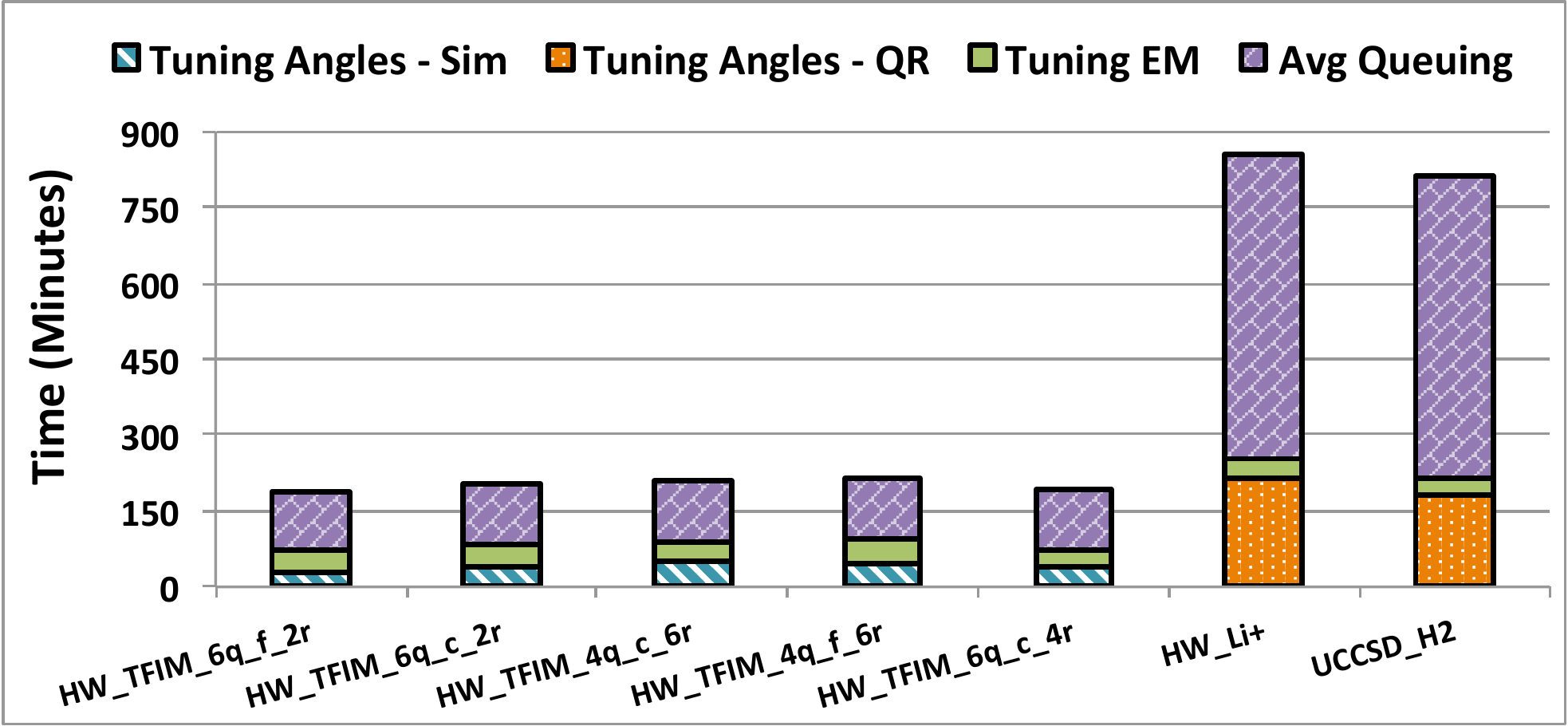}
\caption{Minimal impact of EM tuning on total VQA Execution Time.}
\label{fig:result-3}
\end{figure}

\subsection{Overall Execution Time}
In Fig.\ref{fig:result-3} we compare the total execution time of the different variational approaches.
Time is broken down into 4 components: \circled{1}\ Tuning variational angles in simulation (first 5 applications), \circled{2}\ Tuning variational angles via Qiskit Runtime (molecular chemistry applications), \circled{3}\ Tuning error mitigation via the independent window approach and \circled{4} Queuing time which is the waiting time before accessing quantum machines on the cloud.
Note that the tuning component also includes the actual computation time since computation is part of the tuning loop.

First, note that tuning in simulation is considerably faster than the Qiskit Runtime approach. Note that while this might be the case at the present - it is not a trend that will last. 
This is because: a) simulation will not scale to larger quantum circuits, and b) tuning capabilities of Qiskit Runtime will improve.

Second, queuing times are very large across all executions and are considerably greater than actual runtimes. They are especially high for Qiskit Runtime since: a) only a single machine was available to us which had Runtime capability and b) the machine needs to be held for as many as 5 hours per VQA problem. In order for quantum computing on real machines to scale, especially for long running applications like VQA, it is imperative that there is a greater supply to meet the ever growing demand.

Third, the additional time from error mitigation is under one hour (roughly equals the original tuning time) and is very reasonable for the benefits gained.
In an emerging field like quantum computing, fidelity improvement potentially leading to quantum speedup is worth considerably more than tuning / runtime. 
As circuits scale, tuning times can be reduced by being more selective about which mitigation techniques to apply, where to apply them and how to best tune them.
Further, it is possible that adding additional mitigation parameters can make some of the weaker ansatz gate rotation parameters redundant and might also help break through tuning barren plateaus.
These are worth exploration but are beyond the current scope.

\section{Discussion}

\subsection{Scope of the Current Experiments}
We acknowledge that employing a plethora of other error mitigation techniques (such as those discussed in Section \ref{back-em}), as well as utilizing  optimized ansatz with a potentially more suitable search space, can improve the baseline beyond what we have here. 
But these are orthogonal to the main goal of this paper which is to show that the variational tuning approach can improve existing error mitigation techniques substantially.
We expect that building a more optimal baseline and adding the variational approach to error mitigation on top of it can achieve promising strides in further improving the energy estimates and bring it closer to optimal.
We intend to pursue more optimal baselines in future work.

\subsection{Maximizing the Capability of the Current Approach}
Next, it should be noted that there is potential to further optimize the idle time error mitigation techniques that are tuned in this work.
While we identify some specific features for each technique to variationally tune (the gate positions and the number of DD rounds), the identified features are by no means exhaustive.
For example, spacing between DD sequences could also be tuned to be more optimal than the period / even spacing that is employed in this work.
Further, different DD sequence types (XY, XX, YY etc.) can be employed in conjunction.

\subsection{Variational Approach for other QC Optimizations}
Most importantly, the variational approach introduced in this paper should be decoupled from the specific error mitigation techniques used.
The proposed approach can be utilized to optimally tune a variety of noise-combating and other quantum optimization techniques, both individually and in unison, especially as classical-quantum co-processing frameworks like Qiskit Runtime mature.

We envision the suitability of the variational approach to tuning techniques such as (but not limited to): \circled{1}\ application and machine aware pulse generation for quantum gates~\cite{pulse-gokhale2020optimized, Shi_2019}, \circled{2}\ optimizing native gate sets~\cite{peterson2021optimal}, \circled{3}\ optimizing gate placements~\cite{murali2020software} throughout the circuit , \circled{3}\ trade-offs between different QC optimizations (such as noise-aware mapping~\cite{murali2019noise} and choosing the best qubits for measurement), which as a whole presents a complex tuning space with unknown inter-relations, \circled{4}\ block-based quantum circuit synthesis with block-level variable fidelity thresholds~\cite{wu2021qgo}, and \circled{5}\ possibly even towards better error correction for Fault Tolerant QC.

Further, as shown here, multiple mitigation techniques can always be applied in conjunction through VAQEM if they are suitably integrated into the framework. This is because any destructive interference between techniques will automatically be weeded out by the tuning logic and only the suitable instances of each will be retained.

\begin{figure}[t]
\centering
\fbox{\includegraphics[width=0.98\columnwidth,trim={0cm 0cm 0cm 0cm},clip]{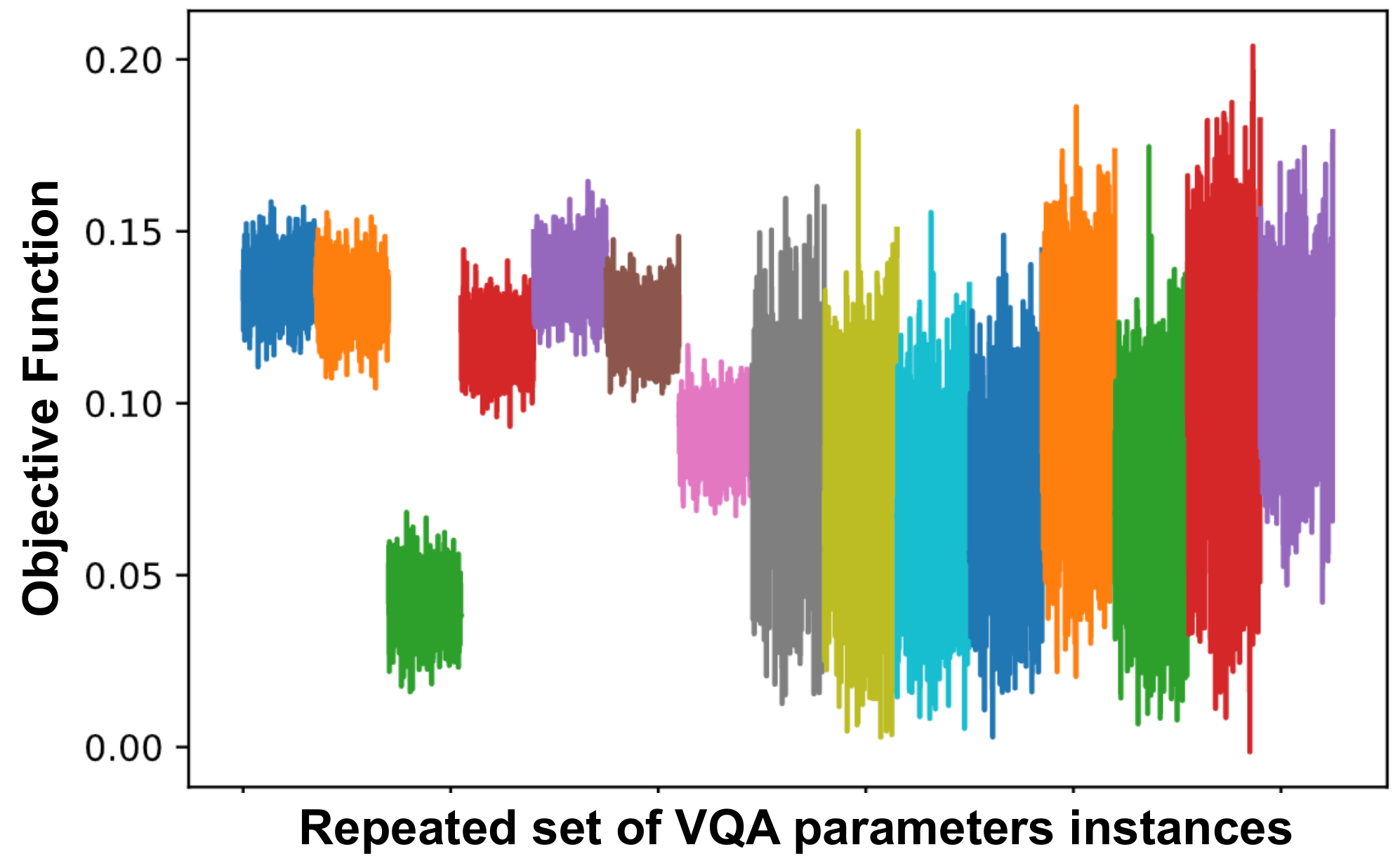}}
\caption{Deviating VQE estimation trends over a 24 hour period on {\tt ibmq\_casablanca}.}
\label{fig:QM}
\end{figure}

\subsection{Temporal Variability in Quantum Machines}
Finally, we discuss the concern of temporal variability in machine characteristics due to re-calibration and machine drift.
While temporal variability impacts all quantum circuit execution in general, it is especially critical to VQA since these are long running jobs and every new iteration of circuit optimization is based on prior measurement feedback from the machine. Thus, stability and continuity in machine characteristics is important.

Fig.\ref{fig:QM} shows how the measured VQA objective function varies for the same set of VQA gate rotation angle parameters over a 24 hour period on {\tt ibmq\_casablanca}. The range of variation is nearly 10-20\% of the ideal objective value, so they are not insignificant.

Each colored cluster is a set of the same 900 VQA parameter configurations added to the ansatz and transpiled in the same manner.
On an ideal machine, it would be expected that the same set of configurations would return the same objective function values over time. 
Clearly in this case the objective function values vary widely over the 24 hour period.
Note that machine re-calibration occurred between the pink and the grey clusters, and the distribution of objective function values vastly changes post calibration. 
Considering long running quantum jobs for VQA and queuing times, it is expected that these VQA problems can often cross machine calibration cycles.
Moreover, even within a calibration cycle, there is considerable variation. 
While most current VQA problems are small enough to be completed in a few hours, and thus less affected by these trends, it is critical to develop tuning approaches which are able to account for these temporal variations.

\section{Conclusion}
While VQAs are highly suited to the NISQ era, the effect of noise is still a significant detriment to VQA's target estimations on real quantum machines - they are far from ideal. 
Thus, it is imperative to employ effective error mitigation strategies in an optimal manner to maximally improve the fidelity of these quantum algorithms on near-term machines.

We proposed VAQEM, which dynamically tailors existing error mitigation techniques to the actual, dynamic noisy execution characteristics of VQAs on a target quantum machine. 
We do so by tuning specific features of these mitigation techniques similar to the traditional rotation angle parameters - by targeting improvements towards a specific objective function which represents the VQA problem at hand. 

While we showed the benefits of VAQEM for two error mitigation techniques in this paper, the proposed variational approach is general and can be extended to many other error mitigation techniques whose specific configurations are hard to select a priori. Integrating more mitigation techniques into the VAQEM framework in the future can lead to formidable gains, potentially realizing practically useful VQA benefits on today's noisy quantum machines.

\section*{Acknowledgement}
This work is funded in part by EPiQC, an NSF Expedition in Computing, under grants CCF-1730082/1730449; in part by STAQ under grant NSF Phy-1818914; in part by DOE grants DE-SC0021526, DE-SC0020289, DE-SC0020331 and DE-SC0020266; and in part by NSF OMA-2016136 and the Q-NEXT DOE NQI Center.
GSR is supported as a Computing Innovation Fellow at the University of Chicago. This material is based upon work supported by the National Science Foundation under Grant \# 2030859 to the Computing Research Association for the CIFellows Project.
KNS is supported by IBM as a Postdoctoral Scholar at the University of Chicago and the Chicago Quantum Exchange.
This research used resources of the Oak Ridge Leadership Computing Facility at the Oak Ridge National Laboratory, which is supported by the Office of Science of the U.S. Department of Energy under Contract No. DE-AC05-00OR22725. 
FTC is Chief Scientist at Super.tech and an advisor to Quantum Circuits, Inc.

\bibliographystyle{IEEEtranS}
\bibliography{refs}

\begin{thebibliography}{10}
\providecommand{\url}[1]{#1}
\csname url@samestyle\endcsname
\providecommand{\newblock}{\relax}
\providecommand{\bibinfo}[2]{#2}
\providecommand{\BIBentrySTDinterwordspacing}{\spaceskip=0pt\relax}
\providecommand{\BIBentryALTinterwordstretchfactor}{4}
\providecommand{\BIBentryALTinterwordspacing}{\spaceskip=\fontdimen2\font plus
\BIBentryALTinterwordstretchfactor\fontdimen3\font minus
  \fontdimen4\font\relax}
\providecommand{\BIBforeignlanguage}[2]{{%
\expandafter\ifx\csname l@#1\endcsname\relax
\typeout{** WARNING: IEEEtranS.bst: No hyphenation pattern has been}%
\typeout{** loaded for the language `#1'. Using the pattern for}%
\typeout{** the default language instead.}%
\else
\language=\csname l@#1\endcsname
\fi
#2}}
\providecommand{\BIBdecl}{\relax}
\BIBdecl

\bibitem{IBMQR}
``{IBM} qiskit runtime,''
  \url{https://quantum-computing.ibm.com/lab/docs/iql/runtime/}.

\bibitem{qiskitruntime}
``Ibm {Q}uantum delivers 120x speedup of quantum workloads with {Q}iskit
  {R}untime,'' \url{https://research.ibm.com/blog/120x-quantum-speedup},
  accessed: 2021-07-30.

\bibitem{IBMQE}
``Ibm quantum experience,'' \url{https://quantum-computing.ibm.com}, accessed:
  2020-11-18.

\bibitem{IBM-SU2}
``{IBM} quantum su2 ansatz,'' \url{https://qiskit.org/documentation/stubs/
  qiskit.circuit.library.EfficientSU2.html}.

\bibitem{IBMQS}
``{IBM} quantum systems,''
  \url{https://quantum-computing.ibm.com/services?systems=all}.

\bibitem{SPSA}
``Spsa: Simultaneous perturbation stochastic approximation method,''
  \url{https://www.jhuapl.edu/spsa/ }.

\bibitem{Qiskit}
H.~Abraham, AduOffei, R.~Agarwal, I.~Y. Akhalwaya, G.~Aleksandrowicz,
  T.~Alexander, M.~Amy, E.~Arbel, Arijit02, A.~Asfaw, A.~Avkhadiev,
  C.~Azaustre, AzizNgoueya, A.~Banerjee, A.~Bansal, P.~Barkoutsos, G.~Barron,
  G.~S. Barron, L.~Bello, Y.~Ben-Haim, D.~Bevenius, A.~Bhobe, L.~S. Bishop,
  C.~Blank, S.~Bolos, S.~Bosch, Brandon, S.~Bravyi, Bryce-Fuller, D.~Bucher,
  A.~Burov, F.~Cabrera, P.~Calpin, L.~Capelluto, J.~Carballo, G.~Carrascal,
  A.~Chen, C.-F. Chen, E.~Chen, J.~C. Chen, R.~Chen, J.~M. Chow, S.~Churchill,
  C.~Claus, C.~Clauss, R.~Cocking, F.~Correa, A.~J. Cross, A.~W. Cross,
  S.~Cross, J.~Cruz-Benito, C.~Culver, A.~D. C{\'o}rcoles-Gonzales, S.~Dague,
  T.~E. Dandachi, M.~Daniels, M.~Dartiailh, DavideFrr, A.~R. Davila,
  A.~Dekusar, D.~Ding, J.~Doi, E.~Drechsler, Drew, E.~Dumitrescu, K.~Dumon,
  I.~Duran, K.~EL-Safty, E.~Eastman, G.~Eberle, P.~Eendebak, D.~Egger,
  M.~Everitt, P.~M. Fern{\'a}ndez, A.~H. Ferrera, R.~Fouilland,
  FranckChevallier, A.~Frisch, A.~Fuhrer, B.~Fuller, M.~GEORGE, J.~Gacon, B.~G.
  Gago, C.~Gambella, J.~M. Gambetta, A.~Gammanpila, L.~Garcia, T.~Garg,
  S.~Garion, A.~Gilliam, A.~Giridharan, J.~Gomez-Mosquera, S.~de~la
  Puente~Gonz{\'a}lez, J.~Gorzinski, I.~Gould, D.~Greenberg, D.~Grinko,
  W.~Guan, J.~A. Gunnels, M.~Haglund, I.~Haide, I.~Hamamura, O.~C. Hamido,
  F.~Harkins, V.~Havlicek, J.~Hellmers, {\L}.~Herok, S.~Hillmich, H.~Horii,
  C.~Howington, S.~Hu, W.~Hu, J.~Huang, R.~Huisman, H.~Imai, T.~Imamichi,
  K.~Ishizaki, R.~Iten, T.~Itoko, JamesSeaward, A.~Javadi, A.~Javadi-Abhari,
  Jessica, M.~Jivrajani, K.~Johns, S.~Johnstun, Jonathan-Shoemaker, V.~K,
  T.~Kachmann, N.~Kanazawa, Kang-Bae, A.~Karazeev, P.~Kassebaum, J.~Kelso,
  S.~King, Knabberjoe, Y.~Kobayashi, A.~Kovyrshin, R.~Krishnakumar,
  V.~Krishnan, K.~Krsulich, P.~Kumkar, G.~Kus, R.~LaRose, E.~Lacal, R.~Lambert,
  J.~Lapeyre, J.~Latone, S.~Lawrence, C.~Lee, G.~Li, D.~Liu, P.~Liu, Y.~Maeng,
  K.~Majmudar, A.~Malyshev, J.~Manela, J.~Marecek, M.~Marques, D.~Maslov,
  D.~Mathews, A.~Matsuo, D.~T. McClure, C.~McGarry, D.~McKay, D.~McPherson,
  S.~Meesala, T.~Metcalfe, M.~Mevissen, A.~Meyer, A.~Mezzacapo, R.~Midha,
  Z.~Minev, A.~Mitchell, N.~Moll, J.~Montanez, M.~D. Mooring, R.~Morales,
  N.~Moran, M.~Motta, MrF, P.~Murali, J.~M{\"u}ggenburg, D.~Nadlinger,
  K.~Nakanishi, G.~Nannicini, P.~Nation, E.~Navarro, Y.~Naveh, S.~W. Neagle,
  P.~Neuweiler, J.~Nicander, P.~Niroula, H.~Norlen, NuoWenLei, L.~J. O'Riordan,
  O.~Ogunbayo, P.~Ollitrault, R.~Otaolea, S.~Oud, D.~Padilha, H.~Paik, S.~Pal,
  Y.~Pang, S.~Perriello, A.~Phan, F.~Piro, M.~Pistoia, C.~Piveteau, P.~Pocreau,
  A.~Pozas-iKerstjens, V.~Prutyanov, D.~Puzzuoli, J.~P{\'e}rez, Quintiii, R.~I.
  Rahman, A.~Raja, N.~Ramagiri, A.~Rao, R.~Raymond, R.~M.-C. Redondo,
  M.~Reuter, J.~Rice, M.~L. Rocca, D.~M. Rodr{\'\i}guez, RohithKarur,
  M.~Rossmannek, M.~Ryu, T.~SAPV, SamFerracin, M.~Sandberg, H.~Sandesara,
  R.~Sapra, H.~Sargsyan, A.~Sarkar, N.~Sathaye, B.~Schmitt, C.~Schnabel,
  Z.~Schoenfeld, T.~L. Scholten, E.~Schoute, J.~Schwarm, I.~F. Sertage,
  K.~Setia, N.~Shammah, Y.~Shi, A.~Silva, A.~Simonetto, N.~Singstock,
  Y.~Siraichi, I.~Sitdikov, S.~Sivarajah, M.~B. Sletfjerding, J.~A. Smolin,
  M.~Soeken, I.~O. Sokolov, I.~Sokolov, SooluThomas, Starfish, D.~Steenken,
  M.~Stypulkoski, S.~Sun, K.~J. Sung, H.~Takahashi, T.~Takawale, I.~Tavernelli,
  C.~Taylor, P.~Taylour, S.~Thomas, M.~Tillet, M.~Tod, M.~Tomasik, E.~de~la
  Torre, K.~Trabing, M.~Treinish, TrishaPe, D.~Tulsi, W.~Turner, Y.~Vaknin,
  C.~R. Valcarce, F.~Varchon, A.~C. Vazquez, V.~Villar, D.~Vogt-Lee,
  C.~Vuillot, J.~Weaver, J.~Weidenfeller, R.~Wieczorek, J.~A. Wildstrom,
  E.~Winston, J.~J. Woehr, S.~Woerner, R.~Woo, C.~J. Wood, R.~Wood, S.~Wood,
  S.~Wood, J.~Wootton, D.~Yeralin, D.~Yonge-Mallo, R.~Young, J.~Yu, C.~Zachow,
  L.~Zdanski, H.~Zhang, C.~Zoufal, Zoufalc, a~kapila, a~matsuo, bcamorrison,
  brandhsn, nick bronn, chlorophyll zz, dekel.meirom, dekelmeirom, dekool,
  dime10, drholmie, dtrenev, ehchen, elfrocampeador, faisaldebouni,
  fanizzamarco, gabrieleagl, gadial, galeinston, georgios ts, gruu, hhorii,
  hykavitha, jagunther, jliu45, jscott2, kanejess, klinvill, krutik2966,
  kurarrr, lerongil, ma5x, merav aharoni, michelle4654, ordmoj, sagar pahwa,
  rmoyard, saswati qiskit, scottkelso, sethmerkel, strickroman, sumitpuri,
  tigerjack, toural, tsura crisaldo, vvilpas, welien, willhbang, yang.luh,
  yotamvakninibm, and M.~{\v{C}}epulkovskis, ``Qiskit: An open-source framework
  for quantum computing,'' 2019.

\bibitem{anschuetz2018variational}
E.~R. Anschuetz, J.~P. Olson, A.~Aspuru-Guzik, and Y.~Cao, ``Variational
  quantum factoring,'' 2018.

\bibitem{biamonte2017quantum}
J.~Biamonte, P.~Wittek, N.~Pancotti, P.~Rebentrost, N.~Wiebe, and S.~Lloyd,
  ``Quantum machine learning,'' \emph{Nature}, vol. 549, no. 7671, pp.
  195--202, 2017.

\bibitem{DDBiercuk_2011}
\BIBentryALTinterwordspacing
M.~J. Biercuk, A.~C. Doherty, and H.~Uys, ``Dynamical decoupling sequence
  construction as a filter-design problem,'' \emph{Journal of Physics B:
  Atomic, Molecular and Optical Physics}, vol.~44, no.~15, p. 154002, Jul 2011.
  [Online]. Available: \url{http://dx.doi.org/10.1088/0953-4075/44/15/154002}
\BIBentrySTDinterwordspacing

\bibitem{bravyi2021mitigating}
S.~Bravyi, S.~Sheldon, A.~Kandala, D.~C. Mckay, and J.~M. Gambetta,
  ``Mitigating measurement errors in multiqubit experiments,'' \emph{Physical
  Review A}, vol. 103, no.~4, p. 042605, 2021.

\bibitem{ding2020systematic}
Y.~Ding, P.~Gokhale, S.~F. Lin, R.~Rines, T.~Propson, and F.~T. Chong,
  ``Systematic crosstalk mitigation for superconducting qubits via
  frequency-aware compilation,'' \emph{arXiv preprint arXiv:2008.09503}, 2020.

\bibitem{farhi2014quantum}
E.~Farhi, J.~Goldstone, and S.~Gutmann, ``A quantum approximate optimization
  algorithm,'' 2014.

\bibitem{giurgica2020digital}
\BIBentryALTinterwordspacing
T.~Giurgica-Tiron, Y.~Hindy, R.~LaRose, A.~Mari, and W.~J. Zeng, ``Digital zero
  noise extrapolation for quantum error mitigation,'' in \emph{2020 IEEE
  International Conference on Quantum Computing and Engineering (QCE)}.\hskip
  1em plus 0.5em minus 0.4em\relax IEEE, 2020, pp. 306--316. [Online].
  Available: \url{http://dx.doi.org/10.1109/QCE49297.2020.00045}
\BIBentrySTDinterwordspacing

\bibitem{Gokhale:2019}
\BIBentryALTinterwordspacing
P.~Gokhale, Y.~Ding, T.~Propson, C.~Winkler, N.~Leung, Y.~Shi, D.~I. Schuster,
  H.~Hoffmann, and F.~T. Chong, ``Partial compilation of variational algorithms
  for noisy intermediate-scale quantum machines,'' \emph{Proceedings of the
  52nd Annual IEEE/ACM International Symposium on Microarchitecture}, Oct 2019.
  [Online]. Available: \url{http://dx.doi.org/10.1145/3352460.3358313}
\BIBentrySTDinterwordspacing

\bibitem{pulse-gokhale2020optimized}
P.~Gokhale, A.~Javadi-Abhari, N.~Earnest, Y.~Shi, and F.~T. Chong, ``Optimized
  quantum compilation for near-term algorithms with openpulse,'' 2020.

\bibitem{Grover96afast}
L.~K. Grover, ``A fast quantum mechanical algorithm for database search,'' in
  \emph{ANNUAL ACM SYMPOSIUM ON THEORY OF COMPUTING}.\hskip 1em plus 0.5em
  minus 0.4em\relax ACM, 1996, pp. 212--219.

\bibitem{hahn}
\BIBentryALTinterwordspacing
E.~L. Hahn, ``Spin echoes,'' \emph{Phys. Rev.}, vol.~80, pp. 580--594, Nov
  1950. [Online]. Available:
  \url{https://link.aps.org/doi/10.1103/PhysRev.80.580}
\BIBentrySTDinterwordspacing

\bibitem{jurcevic2021demonstration}
P.~Jurcevic, A.~Javadi-Abhari, L.~S. Bishop, I.~Lauer, D.~Borgorin, M.~Brink,
  L.~Capelluto, O.~Gunluk, T.~Itoko, N.~Kanazawa \emph{et~al.}, ``Demonstration
  of quantum volume 64 on a superconducting quantum computing system,''
  \emph{Quantum Science and Technology}, 2021.

\bibitem{kandala2017hardware}
A.~Kandala, A.~Mezzacapo, K.~Temme, M.~Takita, M.~Brink, J.~M. Chow, and J.~M.
  Gambetta, ``Hardware-efficient variational quantum eigensolver for small
  molecules and quantum magnets,'' \emph{Nature}, vol. 549, no. 7671, pp.
  242--246, 2017.

\bibitem{DDKhodjasteh_2007}
\BIBentryALTinterwordspacing
K.~Khodjasteh and D.~A. Lidar, ``Performance of deterministic dynamical
  decoupling schemes: Concatenated and periodic pulse sequences,''
  \emph{Physical Review A}, vol.~75, no.~6, Jun 2007. [Online]. Available:
  \url{http://dx.doi.org/10.1103/PhysRevA.75.062310}
\BIBentrySTDinterwordspacing

\bibitem{giurgicatiron2020digital}
\BIBentryALTinterwordspacing
R.~LaRose, A.~Mari, S.~Kaiser, P.~J. Karalekas, A.~A. Alves, P.~Czarnik, M.~E.
  Mandouh, M.~H. Gordon, Y.~Hindy, A.~Robertson, P.~Thakre, N.~Shammah, and
  W.~J. Zeng, ``Mitiq: A software package for error mitigation on noisy quantum
  computers,'' \emph{arXiv:2009.04417}, 2020. [Online]. Available:
  \url{https://arxiv.org/abs/2009.04417}
\BIBentrySTDinterwordspacing

\bibitem{9259985}
W.~Lavrijsen, A.~Tudor, J.~Müller, C.~Iancu, and W.~de~Jong, ``Classical
  optimizers for noisy intermediate-scale quantum devices,'' in \emph{2020 IEEE
  International Conference on Quantum Computing and Engineering (QCE)}, 2020,
  pp. 267--277.

\bibitem{li2017efficient}
\BIBentryALTinterwordspacing
Y.~Li and S.~C. Benjamin, ``Efficient variational quantum simulator
  incorporating active error minimization,'' \emph{Phys. Rev. X}, vol.~7, p.
  021050, Jun 2017. [Online]. Available:
  \url{https://link.aps.org/doi/10.1103/PhysRevX.7.021050}
\BIBentrySTDinterwordspacing

\bibitem{zne4}
\BIBentryALTinterwordspacing
A.~Lowe, M.~H. Gordon, P.~Czarnik, A.~Arrasmith, P.~J. Coles, and L.~Cincio,
  ``Unified approach to data-driven quantum error mitigation,'' \emph{arXiv
  preprint arXiv:2011.01157}, 2020. [Online]. Available:
  \url{https://arxiv.org/abs/2011.01157}
\BIBentrySTDinterwordspacing

\bibitem{mcclean2016theory}
J.~R. McClean, J.~Romero, R.~Babbush, and A.~Aspuru-Guzik, ``The theory of
  variational hybrid quantum-classical algorithms,'' \emph{New Journal of
  Physics}, vol.~18, no.~2, p. 023023, 2016.

\bibitem{moll2018quantum}
N.~Moll, P.~Barkoutsos, L.~S. Bishop, J.~M. Chow, A.~Cross, D.~J. Egger,
  S.~Filipp, A.~Fuhrer, J.~M. Gambetta, M.~Ganzhorn \emph{et~al.}, ``Quantum
  optimization using variational algorithms on near-term quantum devices,''
  \emph{Quantum Science and Technology}, vol.~3, no.~3, p. 030503, 2018.

\bibitem{murali2019noise}
P.~Murali, J.~M. Baker, A.~Javadi-Abhari, F.~T. Chong, and M.~Martonosi,
  ``Noise-adaptive compiler mappings for noisy intermediate-scale quantum
  computers,'' in \emph{Proceedings of the Twenty-Fourth International
  Conference on Architectural Support for Programming Languages and Operating
  Systems}, 2019, pp. 1015--1029.

\bibitem{murali2020software}
P.~Murali, D.~C. McKay, M.~Martonosi, and A.~Javadi-Abhari, ``Software
  mitigation of crosstalk on noisy intermediate-scale quantum computers,'' in
  \emph{Proceedings of the Twenty-Fifth International Conference on
  Architectural Support for Programming Languages and Operating Systems}, 2020,
  pp. 1001--1016.

\bibitem{O_Gorman_2017}
\BIBentryALTinterwordspacing
J.~O’Gorman and E.~T. Campbell, ``Quantum computation with realistic
  magic-state factories,'' \emph{Physical Review A}, vol.~95, no.~3, Mar 2017.
  [Online]. Available: \url{http://dx.doi.org/10.1103/PhysRevA.95.032338}
\BIBentrySTDinterwordspacing

\bibitem{patel2021qraft}
T.~Patel and D.~Tiwari, ``Qraft: reverse your quantum circuit and know the
  correct program output,'' in \emph{Proceedings of the 26th ACM International
  Conference on Architectural Support for Programming Languages and Operating
  Systems}, 2021, pp. 443--455.

\bibitem{peruzzo2014variational}
A.~Peruzzo, J.~McClean, P.~Shadbolt, M.-H. Yung, X.-Q. Zhou, P.~J. Love,
  A.~Aspuru-Guzik, and J.~L. O’brien, ``A variational eigenvalue solver on a
  photonic quantum processor,'' \emph{Nature communications}, vol.~5, p. 4213,
  2014.

\bibitem{peterson2021optimal}
E.~C. Peterson, L.~S. Bishop, and A.~Javadi-Abhari, ``Optimal synthesis into
  fixed xx interactions,'' \emph{arXiv preprint arXiv:2111.02535}, 2021.

\bibitem{pokharel2018demonstration}
B.~Pokharel, N.~Anand, B.~Fortman, and D.~A. Lidar, ``Demonstration of fidelity
  improvement using dynamical decoupling with superconducting qubits,''
  \emph{Physical review letters}, vol. 121, no.~22, p. 220502, 2018.

\bibitem{DDPokharel_2018}
\BIBentryALTinterwordspacing
B.~Pokharel, N.~Anand, B.~Fortman, and D.~A. Lidar, ``Demonstration of fidelity
  improvement using dynamical decoupling with superconducting qubits,''
  \emph{Physical Review Letters}, vol. 121, no.~22, Nov 2018. [Online].
  Available: \url{http://dx.doi.org/10.1103/PhysRevLett.121.220502}
\BIBentrySTDinterwordspacing

\bibitem{preskill2018quantum}
J.~Preskill, ``Quantum computing in the nisq era and beyond,'' \emph{Quantum},
  vol.~2, p.~79, 2018.

\bibitem{romero2018strategies}
J.~Romero, R.~Babbush, J.~R. McClean, C.~Hempel, P.~Love, and A.~Aspuru-Guzik,
  ``Strategies for quantum computing molecular energies using the unitary
  coupled cluster ansatz,'' 2018.

\bibitem{shankar2012principles}
R.~Shankar, \emph{Principles of quantum mechanics}.\hskip 1em plus 0.5em minus
  0.4em\relax Springer Science \& Business Media, 2012.

\bibitem{Sharma_2020}
\BIBentryALTinterwordspacing
K.~Sharma, S.~Khatri, M.~Cerezo, and P.~J. Coles, ``Noise resilience of
  variational quantum compiling,'' \emph{New Journal of Physics}, vol.~22,
  no.~4, p. 043006, Apr 2020. [Online]. Available:
  \url{http://dx.doi.org/10.1088/1367-2630/ab784c}
\BIBentrySTDinterwordspacing

\bibitem{Shi_2019}
\BIBentryALTinterwordspacing
Y.~Shi, N.~Leung, P.~Gokhale, Z.~Rossi, D.~I. Schuster, H.~Hoffmann, and F.~T.
  Chong, ``Optimized compilation of aggregated instructions for realistic
  quantum computers,'' \emph{Proceedings of the Twenty-Fourth International
  Conference on Architectural Support for Programming Languages and Operating
  Systems}, Apr 2019. [Online]. Available:
  \url{http://dx.doi.org/10.1145/3297858.3304018}
\BIBentrySTDinterwordspacing

\bibitem{Shor_1997}
\BIBentryALTinterwordspacing
P.~W. Shor, ``Polynomial-time algorithms for prime factorization and discrete
  logarithms on a quantum computer,'' \emph{SIAM Journal on Computing},
  vol.~26, no.~5, p. 1484–1509, Oct 1997. [Online]. Available:
  \url{http://dx.doi.org/10.1137/S0097539795293172}
\BIBentrySTDinterwordspacing

\bibitem{smith2021error}
K.~N. Smith, G.~S. Ravi, P.~Murali, J.~M. Baker, N.~Earnest, A.~Javadi-Abhari,
  and F.~T. Chong, ``Error mitigation in quantum computers through instruction
  scheduling,'' \emph{arXiv preprint arXiv:2105.01760}, 2021.

\bibitem{souza2012robust}
A.~M. Souza, G.~A. {\'A}lvarez, and D.~Suter, ``Robust dynamical decoupling,''
  \emph{Philosophical Transactions of the Royal Society A: Mathematical,
  Physical and Engineering Sciences}, vol. 370, no. 1976, pp. 4748--4769, 2012.

\bibitem{tannu2019mitigating}
S.~S. Tannu and M.~K. Qureshi, ``Mitigating measurement errors in quantum
  computers by exploiting state-dependent bias,'' in \emph{Proceedings of the
  52nd Annual IEEE/ACM International Symposium on Microarchitecture}, 2019, pp.
  279--290.

\bibitem{tannu2019not}
S.~S. Tannu and M.~K. Qureshi, ``Not all qubits are created equal: a case for
  variability-aware policies for nisq-era quantum computers,'' in
  \emph{Proceedings of the Twenty-Fourth International Conference on
  Architectural Support for Programming Languages and Operating Systems}, 2019,
  pp. 987--999.

\bibitem{temme2017error}
K.~Temme, S.~Bravyi, and J.~M. Gambetta, ``Error mitigation for short-depth
  quantum circuits,'' \emph{Physical review letters}, vol. 119, no.~18, p.
  180509, 2017.

\bibitem{uvarov2020machine}
A.~Uvarov, A.~Kardashin, and J.~D. Biamonte, ``Machine learning phase
  transitions with a quantum processor,'' \emph{Physical Review A}, vol. 102,
  no.~1, p. 012415, 2020.

\bibitem{viola1999dynamical}
L.~Viola, E.~Knill, and S.~Lloyd, ``Dynamical decoupling of open quantum
  systems,'' \emph{Physical Review Letters}, vol.~82, no.~12, p. 2417, 1999.

\bibitem{clops}
A.~Wack, H.~Paik, A.~Javadi-Abhari, P.~Jurcevic, I.~Faro, J.~M. Gambetta, and
  B.~R. Johnson, ``Quality, speed, and scale: three key attributes to measure
  the performance of near-term quantum computers,'' \emph{arXiv preprint
  arXiv:2110.14108}, 2021.

\bibitem{wu2021qgo}
X.-C. Wu, M.~G. Davis, F.~T. Chong, and C.~Iancu, ``{QGo}: Scalable quantum
  circuit optimization using automated synthesis,'' 2021.

\end{thebibliography}

\end{document}